\documentclass[%
 reprint,
 amsmath,amssymb,
 aps,
prb,
]{revtex4-2}

\usepackage{soul}
\usepackage{enumerate}
\usepackage{amsfonts}
\usepackage{bbm}

\usepackage{graphicx}
\usepackage{dcolumn}
\usepackage{bm}
\usepackage{xcolor}
\usepackage[caption=false]{subfig}

\newcommand{\dagga}{{\phantom{\dagger}}}
\newcommand{\bk}{\mathbf{k}}
\newenvironment{eqs}%
{\begin{equation} \begin{aligned}}%
{\end{aligned} \end{equation} }
\newcommand{\beal}{\begin{eqs}}
\newcommand{\eal}{\end{eqs}}
\newcommand{\bealn}{\beal\nonumber}
 \usepackage{braket}

\begin{document}

\preprint{APS/123-QED}

\title{Efficient computational screening of strongly correlated materials - Multi-orbital phenomenology within the ghost Gutzwiller approximation}
\author{Carlos Mejuto-Zaera}
\email{cmejutoz@sissa.it}
\affiliation{International School for Advanced Studies (SISSA), Via Bonomea 265, 34136 Trieste, Italy}
\author{Michele Fabrizio}
\affiliation{International School for Advanced Studies (SISSA), Via Bonomea 265, 34136 Trieste, Italy}

\date{\today}
\begin{abstract}
    The theoretical description of strongly correlated materials relies on the ability to simultaneously capture, on equal footing, the different competing energy scales.
    Unfortunately, existing approaches are either typically extremely computationally demanding, making systematic screenings of correlated materials challenging or are limited to a subset of observables of interest.
    The recently developed ghost Gutzwiller Ansatz (gGut) has shown great promise to remedy this dichotomy.
    It is based on a self-consistency condition around the comparatively simple static one-particle reduced density matrix, yet has been shown to provide accurate static and dynamical observables in one-band systems.
    In this work, we investigate its potential role in the modelling of correlated materials, by applying it to several multi-orbital lattice models.  
    Our results confirm the accuracy at lower computational cost of the gGut, 
    and show promise for its application to materials research.
\end{abstract}
\maketitle

\section{Introduction}

Materials with strong electronic correlations present exciting avenues for device design, owing to the vast palette of different phases of matter they can change between through slight variations of external parameters like temperature or pressure.
The underlying principle enabling this ability to transition between various insulating, metallic, superconducting or magnetic phases is the presence and, crucially, the competition of different energy scales governing the electron dynamics, most notably, the band energy favouring electron delocalization and the Coulomb interaction hampering it.
The theoretical prediction and proposal of correlated materials with targeted properties hinges thus on the ability to accurately model these competing energy scales on equal footing.
This in turn precludes usual mean-field-type models, and the description of strongly correlated materials remains an outstanding challenge in physics.

Arguably the most successful approaches to capture strong correlation in materials are based on the notion of \emph{embedding}.
This refers to mapping the interacting system of interest into an auxiliary impurity model in which the interaction as well as other local effects are inherited by the impurity, while the band energy is represented by the hybridisation with non-interacting degrees of freedom, called \emph{baths}. 
The self-consistency condition required by the embedding procedure entails that the competing energy scales emerge in  
the distribution of the bath parameters, specifically on their energy levels.  
The predominant example of this type of modelling is the dynamical mean-field theory (DMFT)~\cite{Kotliar1996,Kotliar2006}, which has found several successful applications~\cite{Kotliar2001b,Arita2007,Shim2007,Takizawa2009,Haule2010,Park2014,Haule2015,Paul2019}.
Unfortunately, the self-consistency condition in DMFT is formulated around the frequency-dependent one-body Green's function, which is typically hard to approximate accurately since it involves the full spectral structure of the system.
Consequently, using DMFT for materials screening or phase space exploration can be impractical.
While alternative, computationally more inexpensive approximations exist~\cite{deMedici2017}, they can fail to recover the full spectral information.
There is hence a need for a flexible, simplified yet accurate model for correlation in materials.

A recent promising Ansatz in this front is the ghost Gutzwiller (gGut) variational approach~\cite{Lanata2017}.
This is a generalization of the Gutzwiller approximation~\cite{Gutzwiller1963,Gutzwiller1965}, which can be formulated as an embedding strategy.
While in the traditional Gutzwiller method the number of baths is equal to the number of local atomic orbitals, in gGut this is a free parameter, allowing for a more flexible description of competing energy scales.
Moreover, the self-consistency in this approach is formulated in terms of the static one-particle reduced density matrix, and yet it can model full spectral functions accurately~\cite{guerci2019,frank2021,lee2022}.
The gGut Ansatz fulfills thus the conditions of being comparatively computationally inexpensive, while providing with a reliable approximation of static and spectral features of correlations.
So far it has been mainly applied to single-band models, and hence it is necessary to assess its accuracy in multi-orbital systems to determine its actual promise for materials design.
We address this here, and test the gGut approximation on several multi-orbital based phenomenologies, including orbital-selective Mott transitions, Mott-to-band insulator transitions, as well as Hund's metallicity resilient to high interactions.
We find overall excellent performance of the gGut Ansatz and paint a bright perspective for its application in materials applications.

\section{Ghost Gutzwiller Approximation}

\subsection{Review of Gutzwiller Appromiximation}

Before describing the basic structure of the gGut approximation, let us summarize the traditional Gutzwiller Ansatz for lattice models~\cite{Gutzwiller1963,Gutzwiller1965,Bunemann1998,lanata2015,fabrizio2017}. 
We will refer to the Hamiltonian of the system of interest as the physical Hamiltonian $H_{phys}$.
This we will formally split into a non-local, non-interacting contribution $H_{latt}$, depending on the lattice geometry, and all local terms $H_{loc}$

\beal
        H_{phys} &= H_{latt} + \sum_i H_{loc,i}, \\
        H_{latt} &= \sum_{i\neq j,\alpha,\beta}\, t_{i,j}^{\alpha, \beta}\, c^\dagger_{i,\alpha}\,c^\dagga_{j, \beta} = \sum_{\bk, \alpha, \beta} \epsilon_{\bk}^{\alpha, \beta} c^\dagger_{\bk,\alpha}\,c^\dagga_{\bk, \beta}  \\
        H_{loc,i} &= \sum_{\alpha,\beta}\, t_{i,i}^{\alpha,\beta}\,c^\dagger_{i,\alpha}\,c^\dagga_{i,\beta}\\
        &\qquad +\sum_{\alpha,\beta,\gamma,\delta}\, U_{\alpha\beta\,\delta\gamma}\,c^\dagger_{i, \alpha}\,c^\dagger_{i, \beta}\,
        c^\dagga_{i, \delta}\,c^\dagga_{i, \gamma}\,. 
        \label{eq:PhysHamil}
\eal
Here we use the Latin indices ($i,j$) to identify lattice sites, and Greek indices ($\alpha, \beta, \dots$) for physical orbitals, including the spin quantum number $\sigma=\uparrow,\downarrow$, unless specified.
In the second equality of the second line of Eq.~\eqref{eq:PhysHamil}, we have exploited the lattice periodicity to transform from real space orbitals ($i,j$) to the momentum basis $\bk$.

Within the Gutzwiller approximation, one employs the following variational Ansatz~\cite{Yokoyama1987,Capello2005,Lanata2012,fabrizio2017}: starting from an effective single-particle solution to the physical Hamiltonian $\ket{\psi_{qp}}$, e.g. a mean-field solution, recover correlation effects by applying a parametrized projection operator $P$ to it.
The variational degrees of freedom of the approach are thus twofold: the parameters of $P$ as well as those of the effective single-particle solution $\ket{\psi_{qp}}$.
This ``mean-field'' solution is completely determined by an effective single-particle Hamiltonian, which we will refer to as quasi-particle Hamiltonian $H_{qp}$.
As in any mean-field approximation, its role is capturing some of the two-body interaction $U$ through an effective, static one-body Hamiltonian.
In the context of a lattice Hamiltonian with local interactions, this can be done by adding a local one-body potential $\lambda$, as well as by rescaling the non-interacting bands of the physical Hamiltonian $\epsilon_\bk$ with some coefficients $R$.
Hence, one proposes as parametrization for $H_{qp}$ the following expression

\begin{equation}
    H_{qp} = \sum_{\bk,a,b} \left(\sum_{\alpha, \beta}\, R^\dagger_{a,\alpha}\, \epsilon_{\bk}^{\alpha,\beta} \,R_{\beta, b}-\lambda_{a,b}\right)\ d^\dagger_{\bk, a}\,d^\dagga_{\bk, b}.
    \label{eq:qp-Hamil}
\end{equation}

To distinguish the orbitals in the quasiparticle Hamiltonian from the physical ones, we use Latin indices ($a, b$)$\in [1,2,\dots,N_{phys}]$, and denote the creation operators by $d$ instead of $c$. 
In the parametrization of Eq.~\eqref{eq:qp-Hamil}, a direct connection can be drawn between the rescaling factors $R^\dagger,\ R$ and the quasiparticle renormalization factors $Z$.

The projection operator $P$ is defined, in general, as a linear map between the Hilbert spaces of $H_{qp}$ and $H_{phys}$.
Therefore, it involves a number of variational parameters which grows exponentially in the number of particles. 
This is reduced somewhat in the case of strictly local interactions, by decomposing $P$ into a product of identical local projectors.
Still, an exponential scaling in the number of quasiparticle and physical orbitals remains, which would result in an extremely tedious optimization procedure.
This explicit optimization can be avoided in the infinite dimensional limit, where expectation values over the Gutzwiller Ansatz wave function $\ket{\psi_0}=P\ket{\psi_{qp}}$ are equivalent to expectation values over wave functions of a local impurity model $H_{imp}$ defined over both the physical and quasiparticle orbitals~\cite{lanata2015}.
The larger number of orbitals ensures that the wave functions of the impurity model have the same number of degrees of freedom as the operator $P$.
The variational optimization of the parameters in $P$ turns thus into finding the ground state wave function of $H_{imp}$, defined as

\beal
    H_{imp} &= H_{loc} + \sum_{\alpha, a}\,\Big(V_{\alpha, a} d^\dagger_a\, c^\dagga_\alpha + \mathrm{h.c.}\Big)\\
    & \qquad -\sum_{a,b}\, \lambda_{a,b}^c\, d^\dagger_{a}\,d^\dagga_b.
    \label{eq:imp-H}
\eal

Here, $H_{loc}$ corresponds to the local interaction term in the physical Hamiltonian $H_{phys}$, for a single lattice site.
As anticipated, $H_{imp}$ is a Hamiltonian with $2 N_{phys}$ orbitals, involving at the same time the physical ($c_\alpha$) and the quasiparticle states ($d_a$).
The quasiparticle orbitals have their own effective one-body potential $\lambda^c$, related to $\lambda$ in $H_{qp}$, and are coupled to the physical orbitals through the hybridization $V$, which takes a similar role as $R$.
Finally, the $\lambda$ and $R$ parameters need to be fixed, 
and this is done through a self-consistency condition, involving the one-body reduced density matrices (1-RDM) of the impurity and quasiparticle ground states

\begin{equation}
    \langle\, d^\dagga_b\,d^\dagger_a\,\rangle_{imp} \overset{!}{=} \frac{1}{V_{BZ}}\,\sum_\bk\;\langle\, d^\dagger_{\bk, a}\, d^\dagga_{\bk, b} \,\rangle_{qp}\,.
    \label{eq:scf}
\end{equation}

where $V_{BZ}$ is the Brillouin zone volume. 
Taking everything into account, this substitutes the original variational optimization with an exponentially large number of degrees of freedom by a self-consistently coupled pair of impurity and non-interacting lattice problems, within an infinite dimensional approximation. 

Despite its wide application for describing strong correlation in extended and molecular systems alike~\cite{Deng2008,Wang2010,Yao2012,Lanata2013,Lanata2014,Borghi2014,lanata2015,Tian2015,Peng2021,Ye2022}, the Gutzwiller approximation has some clear shortcomings.
In the example of the Mott transition in a single band Hubbard model at half-filling, the Gutzwiller Ansatz describes the onset of insulating behaviour at large $U/t$ in terms of a divergent effective electronic mass~\cite{Brinkman1970}.
While this satisfactorily accounts for the narrowing and eventual vanishing of the coherent metallic band at the Fermi level, it fails to describe the incoherent high energy Hubbard bands, which characterize the nature of the insulating state and relate it to the atomic limit~\cite{Fulde1995_gutz}.
Essentially, the Gutzwiller approach fails here to describe the competing low and high energy features of the model simultaneously.
In a multi-orbital setting, where this competition becomes key to understanding the different possible phases, this is a serious limitation.
The ghost Gutzwiller formulation addresses precisely this matter. 

\subsection{Heuristic of ghost Gutzwiller}

Starting from the above scheme for the Gutzwiller approximation, the ghost Gutzwiller (gGut) extension increases the variational flexibility by including $N_g$ additional orbitals, often referred to as \emph{ghosts}, to the quasiparticle (and hence impurity) Hamiltonian~\cite{Lanata2017,guerci2019,frank2021,lee2022}. 
From a variational standpoint, there is no need to restrict the number of quasiparticle orbitals to be $N_{phys}$, as is done in standard Gutzwiller.
Indeed, the quasiparticle orbital index $a$ can in principle run from $1$ to $N_{eff} > N_{phys}$, as long as the subsequent projector operator $P$ restricts the Ansatz wave function to the physical orbitals. 

Besides pure optimizational flexibility, this generalization resolves the problem of describing multiple competing energy scales simultaneously:
The additional auxiliary quasiparticle orbitals take the role of capturing the different energy scales that are missing in the regular Gutzwiller approximation.
They fulfill essentially the same role as that of bath orbitals in embedding models of electronic correlation~\cite{Caffarel1994,Kotliar1996,Koch2008,Liebsch2011,Knizia2012,Knizia2013}, and hence we will refer to them also as \emph{baths}.
In this light, the limitations of the standard Gutzwiller approach when describing the Mott insulating phase in the single-band Hubbard model at half-filling becomes intuitive to understand: the single quasiparticle degree of freedom (bath) in that case cannot possibly account simultaneously for both Hubbard bands at $\pm U/2$ in the insulating phase as well as the coherent peak at zero energy in the metallic phase.
But just adding two additional auxiliary degrees of freedom, for a total of three ($N_{eff}=N_{phys}+N_{g}=1+2$) quasiparticle orbitals, it becomes possible to model both the metallic band, already captured in regular Gutzwiller, and the Hubbard bands.
In this way, the ghost Gutzwiller Ansatz can account for a homogeneous spectral description of the full metal-insulator Mott transition~\cite{Lanata2017,frank2021,guerci2019,lee2022}.
Further increasing the number of baths allows then to account for more competing energy scales, making this generalization particularly attractive for multi-orbital models and \emph{ab initio} simulations.

\subsection{Procedural Formulation}

In this subsection, we discuss the central equations of the gGut formalism, motivating their physical significance and providing a minimal prescription for its implementation.
Some more specialized details are collected in the appendices.
We refer the interested reader to the existing literature for a formal derivation of these equations~\cite{lanata2015,Lanata2017}.

As described heuristically above, gGut can be formulated as a self-consistent problem relating a non-interacting lattice Hamiltonian $H_{qp}$ with an interacting finite size impurity Hamiltonian $H_{imp}$ through their respective 1-RDMs.
The input for the approximation is a system Hamiltonian of interest $H_{phys}$, and the number of auxiliary degrees of freedom $N_g$ to be included in $H_{qp}$ and $H_{imp}$.
Upon convergence, the output of the calculation are the quasiparticle renormalizations $R$ and potential $\lambda$ which define $H_{qp}$ (cf. Eq.~\eqref{eq:qp-Hamil}).
One can then evaluate quantities of interest of $H_{qp}$, such as one- and two-body correlators, and thereby estimate the corresponding properties of $H_{phys}$ upon projecting with $R$.
While this suggests a formulation of gGut around the self-consistent convergence of variationally optimal $R$ and $\lambda$, it is computationally advantageous to instead consider the convergence for $R$ and the quasi-particle 1-RDM $\Delta_{ab} = \langle\, d^\dagger_{a}\, d^\dagga_{b} \,\rangle_{qp}$.

At the $\ell$-th iteration, having a pair of parameters $R^{\ell}$ and $\Delta^\ell$, the first step in the gGut approach is finding the quasiparticle potential $\lambda^{\ell}$ that, for the given renormalization factors $R^\ell$, generates the current target 1-RDM $\Delta^{\ell}$. 
In other words, to enforce the self-consistency condition in Eq.~\eqref{eq:scf_cond}, one finds the $\lambda^{\ell}$ such that the ground-state $\ket{\psi_{qp}}$ of $H_{qp}$ fulfills
\begin{equation}
    \frac{1}{V_{BZ}}\,\sum_\bk\,\braket{\psi_{qp}|d^\dagger_{\bk,a}\, d^\dagga_{\bk,b}|\psi_{qp}}=\Delta^{\ell}_{a,b}.
    \label{eq:lambd_eq}
\end{equation}

This represents an optimization problem, essentially a fit for the components of $\lambda^{\ell}$. Details on its implementation are given in the appendix. 

Next, having $R^{\ell}$ and $\lambda^{\ell}$, we can evaluate the terms defining the impurity Hamiltonian. 
We start with the hybridizations $V^\ell$, which describe the coupling of the local (physical) impurity orbitals in $H_{imp}$ to the bath.
Since the bath essentially substitutes the rest of the lattice, $V^\ell$ needs to capture the effective kinetic energy of electrons leaving the local unit cell, including approximate correlation effects due to the interactions with electrons in the other unit cells.
Consistent with this intuition, the formal derivation of the method leads to the following expression for $V^\ell$ at the $\ell$-th iteration, 
\begin{equation}
    \sqrt{\Delta^{\ell}\big(\mathbb{I}-\Delta^{\ell}\big)\;}\cdot V^\ell = \frac{1}{V_{BZ}}\,\sum_{\bk} \Delta^{\ell,t}_\bk \cdot R^\ell{^\dagger} \cdot \epsilon_{\bk},
    \label{eq:V_eq}
\end{equation} 

where $\cdot$ marks matrix products, $\mathbb{I}$ is the identity matrix, and $\Delta^{\ell,t}_\bk$ is the transpose of the quasiparticle 1-RDM in $\bk$-space, ie. $\Delta^{\ell,t}_{\bk,a,b} = \braket{\psi_{qp}|d^\dagger_{\bk,b}\,d^\dagga_{\bk,a}|\psi_{qp}}$. 
We observe that the RHS of the equation does indeed contain the effective lattice (quasiparticle) kinetic energy, half-projected into the physical degrees of freedom with $R^\ell$.
Note that, consistently with Eq.~\eqref{eq:imp-H}, $V^\ell$ is a $\left(N_{eff}\times N_{phys}\right)$ matrix.
Further, since $\Delta^\ell$ has eigenvalues between 0 and 1, the radical of the square-root is always positive semi-definite.

Next follows the evaluation of $\lambda^\ell{^c}$, the bath potential in the impurity Hamiltonian. 
This term should model the effective lattice correlations, and hence has to be directly related to the quasiparticle potential $\lambda^\ell$ at the same iteration.
Indeed, it equals $\lambda^\ell$ up to a correction term due to the feedback coupling between the local impurity and the rest of the lattice, involving the renormalization $R$, hybridization $V$ and the following matrix derivatives

\beal
    &\lambda^c{^\ell}_{a,b} = -\lambda^\ell_{a,b}\\
    &\quad +\left\{\frac{\partial}{\partial \Delta^\ell_{a,b}}
    \left[R^\ell\cdot \sqrt{\Delta^{\ell}(\mathbb{I}-\Delta^{\ell})\;}\cdot V^\ell \right]+\mathrm{h.c.}\right\}\,.
    \label{eq:lambd_c}
\eal

Note that the derivative only acts on the square root over $\Delta^\ell (\mathbb{I}-\Delta^{\ell})$.

Having the $V^\ell$ and $\lambda^c{^\ell}$ matrices, the impurity Hamiltonian $H_{imp}$ is fully defined, and its ground state 1-RDM $\Delta^\ell_{imp}$ can be determined numerically. 
This can be done in principle, with any impurity solver, as long as one obtains a faithful representation of $\Delta^\ell_{imp}$. 
When practical, exact diagonalization (ED) can be used~\cite{Caffarel1994,Amaricci2022}, and otherwise truncated approximations such as configuration interaction (CI)~\cite{Zgid2011,Zgid2012,Lu2014,Go2015,Go2017,Mejuto2019,WilliamsYoung2023,Werner2023} or tensor network approaches~\cite{Garcia2004,Nishimoto2004,Peters2011,Wolf2014a,Wolf2014b,Wolf2015a,Wolf2015b,Bauernfeind2017,Paeckel2019} can be readily implemented. 
In this work, we use ED. 

Finally, we can close the self-consistent cycle by evaluating a new quasipartcile renormalization $R^{\ell+1}$ and quasiparticle 1-RDM $\Delta^{\ell+1}$ from the impurity 1-RDM as:
\begin{equation}
    \begin{split}
        \Delta^{\ell+1} &= \mathbb{I} - \Delta^\ell_{imp,bath-bath}, \\
        R^{\ell+1}\cdot\sqrt{\Delta^{\ell+1}(\mathbb{I}-\Delta^{\ell+1})} &= \Delta^{\ell,t}_{imp, bath-imp}.
    \end{split}
    \label{eq:scf_cond}
\end{equation}

Here, $\Delta^\ell_{imp, bath-bath}$ refers to the $\ell$-th step impurity 1-RDM restricted to the bath degrees of freedom, whereas $\Delta^\ell_{imp, bath-imp}$ is its off-diagonal component mixing bath and impurity orbitals.

The described iteration is repeated until either the matrices $(R, \Delta)$ converge within some threshold, or until the variational lattice energy does.
This energy is evaluated at each iteration as the sum of the energy of the projected quasiparticle lattice and the local energy of the impurity model

\begin{equation}
    E_{var} = \braket{\psi_{imp}|H_{loc}|\psi_{imp}}+\braket{\psi_{qp}|R\cdot H_{qp}\cdot R^\dagger|\psi_{qp}}.
    \label{eq:varE}
\end{equation}

Before turning to the reliability of the gGut approximation to multi-orbital contexts, a brief discussion on its similarities and differences with dynamical mean-field theory follows in the next subsection.
We just note in passing that Eq.~\eqref{eq:V_eq} and the second equation in Eq.~\eqref{eq:scf_cond} are ill-defined if $\Delta^\ell$ or $\Delta^{\ell+1}$ have eigenvalues of exactly 0 or 1.
This may happen because some of the effective orbitals represent fully uncorrelated bands, cf. the band insulator phase at high crystal-field splitting in the results section.
In this case, the iteration can be stabilized by adding some artificial one-body coupling terms into the Hamiltonian, to enforce a finite degree of partial occupation in all bands.

\subsection{Parallels with Dynamical Mean-field Theory}

In the impurity model based formulation presented above, the gGut approximation shares many similarities with a different non-perturbative approach to describe strong electronic correlation: the dynamical mean-field theory (DMFT), particularly in its Hamiltonian formulation.
In this subsection, we will briefly discuss these similarities, as well as point out the main differences between these complementary frameworks.

Perhaps the most obvious similarity between the two is the central role of the infinite dimensional limit.
As mentioned above, this is invoked in the gGut formalism to allow for the evaluation of expectation values with the Gutzwiller Ansatz $P\ket{\psi_{qp}}$ in terms of a completely local impurity model.
This infinite dimensional limit underlies the formulation of DMFT as well, but there is a key difference to be acknowledged:
DMFT represents, in its single site formulation, the exact solution of the many-body problem in the infinite dimensional limit, whereas the gGut method in its impurity model formalism is the exact solution \emph{of a variational wavefunction}.
However, this distinction is in no way indicative that gGut is inferior in general terms to DMFT, particularly since both find their applications predominantly in finite dimensional systems, where both are approximate.
In this equal ground, gGut carries the advantage of being (approximately) variational in the ground state energy, and hence offering a more transparent way of choosing between competing fix-points of the self-consistency.
In both cases, spatially non-local approximations are only captured at a single-particle level.

From a modelling perspective, both Hamiltonian based DMFT and gGut operate in a completely analogous way:
They substitute a fully interacting lattice by a pair of an interacting impurity model and non-interacting lattice coupled via a self-consistency condition.
The main steps in these self-consistencies are in both cases the solution of the impurity problem, and a fit of the parameters of the non-interacting lattice model.
The impurity problem is significantly simpler in the gGut case, since the quantity of interest is the static 1-RDM, which is less computationally expensive to evaluate and approximate than the frequency dependent Green's function at the core of DMFT.
In this regard, gGut is reminiscent of the density matrix embedding theory (DMET) approach~\cite{Knizia2012,Knizia2013}.
Surprisingly, despite being formulated around the static 1-RDM, gGut can provide qualitatively excellent Green's functions in great agreement with DMFT~\cite{Lanata2017,guerci2019,frank2021,lee2022}.
This marks gGut as particularly attractive to treat multi-orbital systems and even \emph{ab initio} models of materials, which can be more challenging to access with DMFT.

Although the impurity problem in gGut is relatively simpler than in DMFT, this is not the case for the fitting step.
As discussed in the appendix, the momentum summation that is intrinsic in the optimization problem in gGut makes it considerably harder than the fitting step in DMFT.
Moreover, while numerical strategies to simplify the fitting in DMFT have been proposed~\cite{Mejuto2020}, it is not obvious that these can be applied in the gGut setting.
Still, from a computational complexity scaling point of view, the main bottleneck is the impurity solver, and hence the simplifications presented by gGut in this regard outweigh the increased complications of the fit.

Taking everything into account, gGut presents a flexible and qualitatively accurate approximation method to study the spectral features of strongly correlated systems.
Its reduced computational scaling, when compared with established methods as DMFT, identifies it as an ideal candidate to perform extensive phase space explorations of multi-orbital models and \emph{ab initio} systems alike.
In the following section, we present exemplary results for investigations of that kind in 2- and 3-band models with markedly different phenomenologies.

\section{Multi-Orbital Phenomenology with ghost Gutzwiller}

To investigate the reliability of the gGut approximation for the description of strong correlation in multi-orbital systems, we concentrate on three different paradigmatic and well-established phenomenologies.
These include

\begin{itemize}
    \item Orbital-selective Mott transitions on a 2-band Hubbard model with orbital-dependent bandwidths $D_i$, $i=1,2$ (cf. Refs.~\cite{deMedici2005,Ferrero2005}).
    \item Mott insulator to band insulator transition in 2-band Hubbard model with Hund coupling $J$ and crystal field splitting $\Delta$ (cf. Ref.~\cite{Werner2007}).
    \item Mott vs. Hund insulating behaviour, with an intermediate Hund metallic phase surviving high interaction strengths on a 3-band Hubbard model with Kanamori interaction (cf. Ref.~\cite{Isidori2019}).
\end{itemize}

The competing energy scales in these models are represented by the local Hubbard repulsion $U$, the  bandwidths $D_i$, the crystal field terms $\Delta$ and the Hund coupling $J$.
The one-body terms $D_i$ and $\Delta$ allow for more realistic, i.e. materials-like, model Hamiltonians, which effectively capture the inequivalent spatial overlap of different local orbitals, and the coupling of electronic motion to lattice distortions respectively.
The Hubbard repulsion $U$ allows for the minimal description of electronically driven metal-insulator transitions, bridging between the delocalized and atomic limits~\cite{Imada1998}.
Finally, the Hund coupling $J$ accounts for multiplet splitting in the atomic electronic configuration, and has been shown to be central to the description of the electronic properties in Fe-based superconductors~\cite{Werner2008,Haule2009,Hansmann2010,deMedici2014,Fanfarillo2015,Hoshino2016,deMedici2017,Fanfarillo2017,VillarArribi2018,VillarArribi2021}.

In all simulations, we set the number of effective orbitals $N_{eff}$ equal to three times the number of physical orbitals $N_{phys}$, which as discussed above is the minimal ratio to describe the Mott transition accurately within gGut.
The impurity models in the 2 and 3 band models have thus 8 and 12 orbitals respectively, and are solved with exact diagonalization (ED).
Further, unless otherwise specified, a Bethe lattice to simplify the momentum summation is assumed.

\subsection{Orbital-Selective Mott Transition}

\begin{figure}
    \centering
    \includegraphics[width=0.5\textwidth]{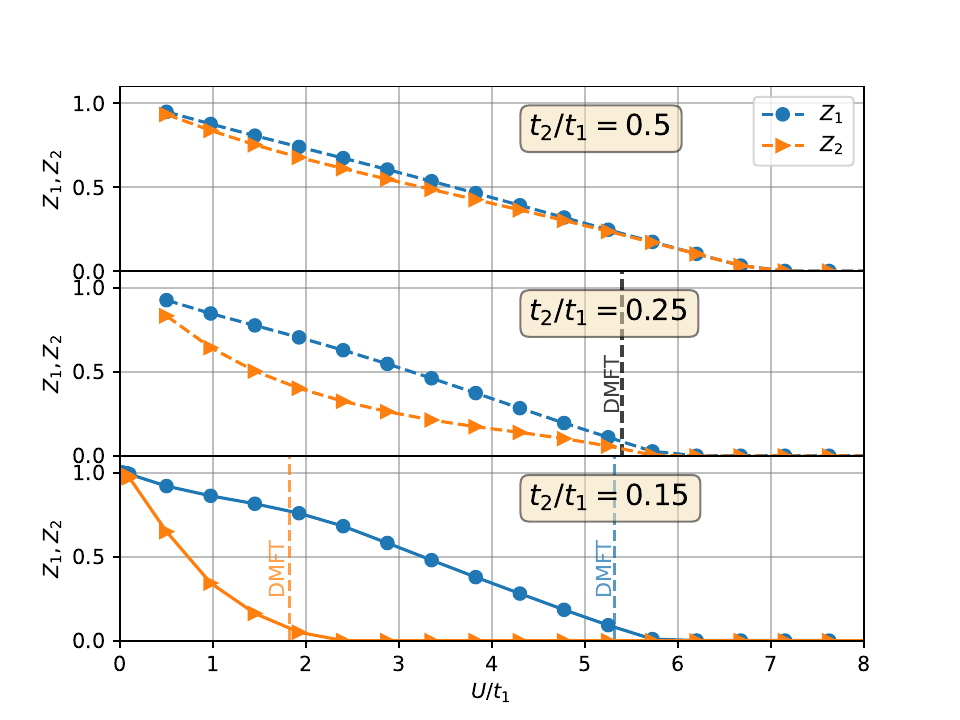}
    \caption{Quasiparticle renormalization factors $Z_i$ for the two-band model in Eq.~\eqref{eq:2bandOSMT}, for three different bandwidth ratios $t_2/t_1$. An orbital selective Mott transition is observed for the largest bandwith difference $t_2/t_1 = 0.15$. Where available, the critical $U/t$ from DMFT are shown as vertical lines, taken from Ref.~\cite{deMedici2005}.}
    \label{fig:OSMT}
\end{figure}

The first immediate complication when moving away from the single-band picture of the Mott transition concerns the fact that, in realistic systems, multiple bands are involved with potentially different nature.
This can be represented by different associated bandwidths, or even on-site energies (c.f. next subsection).
In the former case, this can lead to orbital-selective metal insulator transitions, corresponding to systems where some of the electrons are itinerant (metallic), while others are localized (insulating).
This possibility attracted special interest after Ca$_{2-x}$Sr$_x$RuO$_4$ was proposed to exhibit exactly this behavior~\cite{Anisimov2002}, which was then eventually shown to be theoretically possible using different models~\cite{Liebsch2003,Koga2004,deMedici2005,Ferrero2005}.
It is therefore of interest to investigate the description of orbital-selective Mott transitions within the gGut framework.
We focus here on the two band Hubbard model in Ref.~\cite{deMedici2005}.
The local Hamiltonian here reads
\beal
    &H_{loc,i} = -\mu\sum_{a,\sigma}\, n_{ia\sigma}+U\sum_a\, n_{ia\uparrow}\, n_{ia\downarrow}  \\
    &\qquad+ (U-2J)\, \sum_{\sigma}\,n_{i1\sigma}\,n_{i2\bar\sigma}\\  &\quad +(U-3J)\,\sum_{\sigma}\,n_{i1\sigma}\,n_{i2\sigma}\\
    &\;-J\left[c^\dagger_{i1\uparrow}\,c^\dagga_{i1\downarrow}\,c^\dagger_{i2\downarrow}\,c^\dagga_{i2\uparrow}+c^\dagger_{i1\uparrow}\,c^\dagger_{i1\downarrow}\,c^\dagga_{i2\uparrow}\,c^\dagga_{i2\downarrow}+\mathrm{h.c.}\right]\,,
\label{eq:2bandOSMT}
\eal
with an orbital dependent hopping amplitude $t_a$ in the lattice Hamiltonian implied.
We perform gGut simulations with six baths (four ghosts), at half-filling with $J = 0$ and different ratios $t_2/t_1 = 0.5, 0.25, 0.15$.
Fig.~\ref{fig:OSMT} shows the resulting quasiparticle renormalization factors $Z_a$ for both bands as a function of the local Coulomb repulsion strength $U/t_1$.
As can be seen already for $t_2/t_1 = 0.25$, gGut captures the markedly different behavior of the quasiparticle renormalization between both bands, driven by the difference in their bandwidths.
Upon further decreasing the hopping amplitude ratio the gGut approximation is also able to satisfactorily recover the expected orbital-selective Mott transition, presenting a regime of phase space where one of the two bands is metallic while the other becomes Mott insulating.
The onset of the orbital-selective Mott transition as well as the overall values for $Z_a$ as a function of $U/t_1$ are in good agreement with the DMFT calculations in Ref.~\cite{deMedici2005}. 
The critical interaction strength at which the metal to insulator transition occurs is slightly overestimated in the gGut description, which is reasonable given its larger mean-field character compared to DMFT.

\subsection{Mott Insulator to Band Insulator Transition}

\begin{figure}
    \centering
    \includegraphics[width=0.5\textwidth]{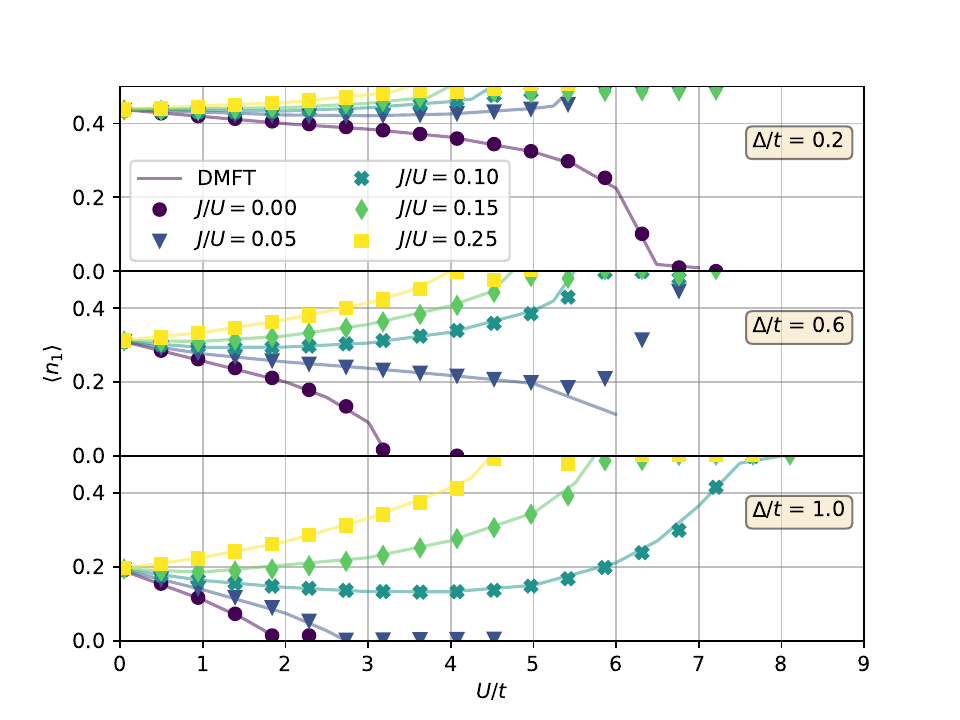}
    \caption{Average charge density in the first band of the two-band model with crystal-field splitting of Eq.~\eqref{eq:2bandCF}, for different parameter regimes. Values of the crystal-field splitting $\Delta/t = 0.2, 0.6, 1.0$ are presented in different subfigures. For each crystal-field splitting, different $\langle n_1\rangle$ vs $U/t$ are shown for different values of $J/U$, in a gradated color scheme with various markers. A clear difference is seen between band insulators (for which $\langle n_1\rangle\rightarrow0$) and Mott insulators (where $\langle n_1\rangle\rightarrow0.5$). For comparison, results from CTQMC-DMFT from Ref.~\cite{Werner2007} are presented as solid lines.}
    \label{fig:i2i}
\end{figure}

\begin{figure*}
    \centering
    \includegraphics{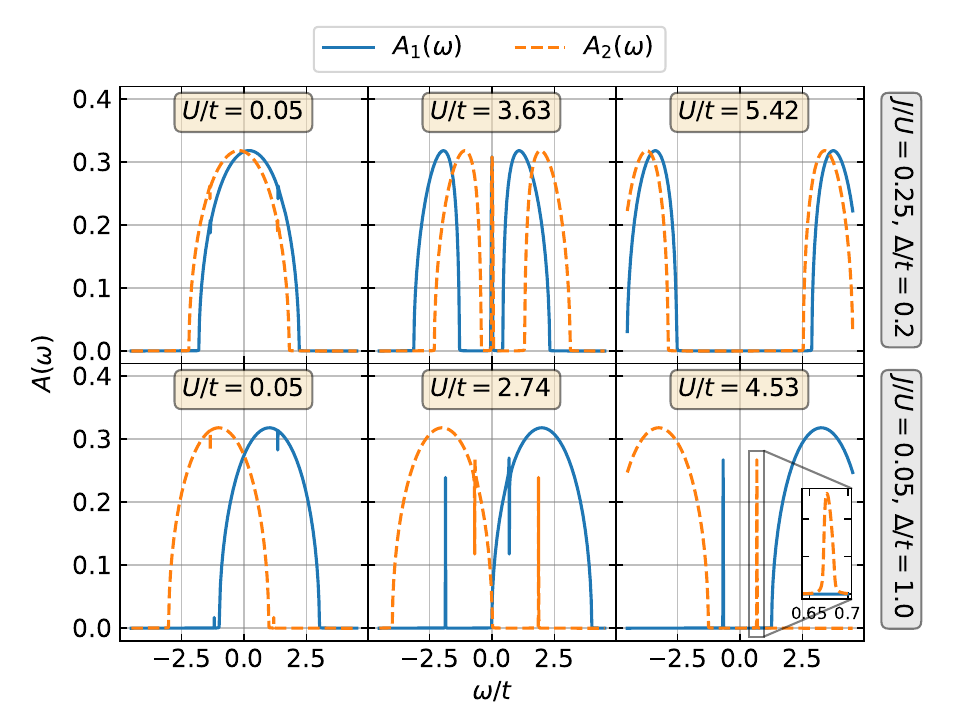}
    \caption{Orbital resolved spectral functions $A_i(\omega) = -\frac{1}{\pi}\Im G_{ii}(\omega)$ for the two-band model with crystal-field splitting of Eq.~\eqref{eq:2bandCF}. Results are shown for the Mott-regime (high $J/U$, small $\Delta/t$, upper panels), and the band regime (low $J/U$, large $\Delta/t$, lower panels), for different values of $U/t$. The inset in the lower right panel shows that all features have a finite width.}
    \label{fig:M2B_Aws}
\end{figure*}

Insulator to insulator transitions are a common and central motif in the understanding of electronically driven lattice geometry transitions in several materials, such as ferroelectric perovskites~\cite{Fabrizio1999} and organic charge-transfer compounds~\cite{Nagaosa1986,Tinkani2009}.
This becomes particularly relevant in multi-orbital models, where lattice distortions can lead to asymmetric crystal fields breaking orbital degeneracy.
In this spirit, we study crystal-field driven Mott-to-band insulator transitions in the two-band Hubbard-Kanamori model in Ref.~\cite{Werner2007}.
The local Hamiltonian in this case follows
\beal
    \widetilde{H}_{loc,i} &=
    H_{loc,i}  + \Delta\,\sum_{\sigma}\,\big(n_{i1\sigma}-n_{i2\sigma}\big)\,,
    \label{eq:2bandCF}
\eal
where $H_{loc,i}$ is defined in Eq.~\eqref{eq:2bandOSMT}, and 
we have introduced a crystal field splitting term $\Delta$.
We perform gGut simulations with six baths (four ghosts) at half-filling for different ratios of $\Delta/t$ and $J/U$.
The final local orbital occupation of the upper band $\langle\, n_{1\sigma}\,\rangle$ is shown in Fig.~\ref{fig:i2i}.
For large crystal-field splitting and moderate $J/U$, the local orbital occupation of the upper band goes to zero at large $U/t$.
By particle-hole symmetry, the other band is completely full in this scenario, and hence the system essentially becomes a simple band insulator~\footnote{It is worth noting that in this band insulator regime we encounter the problem of singular $\sqrt{\Delta(\mathbb{I}-\Delta)\;}$ in Eq.~\eqref{eq:V_eq} and~\eqref{eq:scf}. The band insulator phase can be nonetheless stabilized within the gGut approximation by adding a nearest-neighbor hopping coupling the lower and upper bands in the lattice Hamiltonian. This prevents the bands from completely filling up (emptying) while still allowing the insulating gap to open.}.
At smaller crystal field splitting, or sufficiently large Hund's coupling, the local occupation of the upper band goes instead towards 0.5, and similarly so for the lower band.
This corresponds to a conventional Mott insulator.

The differences between the band and Mott insulator become especially apparent when examining the orbital resolved spectral functions $A_i(\omega) = -\frac{1}{\pi}\Im  G_{ii}(\omega)$, which we present in Fig.~\ref{fig:M2B_Aws}.
In the ``Mott regime'' shown in the upper panels of Fig.~\ref{fig:M2B_Aws}, i.e. for large Hund interaction and small crystal-field splitting $\Delta$, the spectral function develops in the usual Mott-Hubbard way as a function of $U/t$:
At small interactions the full $A(\omega) = \sum_iA_i(\omega)$ is formed by a single metallic dome at the Fermi level, composed of single domes for each orbital shifted by $\Delta$.
Increasing the interaction strength, the metallic band at the Fermi level narrows, while high-energy Hubbard bands form.
Noticeably, these Hubbard bands are also composed from contributions from both orbitals, shifted from each other.
Finally, after some critical interaction, the metallic feature at $\omega = 0$ vanishes, and only the Hubbard bands remain.
Then ``band regime'', corresponding to small Hund interactions and large crystal-field splittings, is shown in the lower panels of Fig.~\ref{fig:M2B_Aws}.
For small Hubbard interactions, the situation is similar to the one found in the ``Mott regime'', except for a larger splitting between the bands of each orbital.
However, increasing the Hubbard interaction strength does not generate the appearance of high-energy Hubbard bands, but instead just increases the shift between the two bands, until they eventually completely separate, resulting in an insulating gap.
This is clearly a simple, one-body band insulator, in contrast with the Mott insulator shown in the upper panels of Fig.~\ref{fig:M2B_Aws}.

The gGut description predicts this transition in excellent agreement with DMFT results using a continuous time quantum Monte Carlo solver~\cite{Werner2007}, which notably does not discretize the bath.
In fact, the local occupation numbers in Fig.~\ref{fig:i2i} agree in this case to a quantitative level with the DMFT results.
This is partly because static properties, such as occupation numbers, are significantly easier to capture than dynamical quantities, such as the renormalization factor $Z$.
Nonetheless, this level of agreement shows the immense value of the gGut approximation as a reliable and yet computationally inexpensive variant to explore strongly correlated phenomena in the phase space of multi-orbital systems.

In between the Mott and band insulating phases of this 2-band model with crystal field splitting, the system remains metallic.
Still, for any field strength $\Delta/t$ and Hund coupling $J/U$ there is some Hubbard repulsion $U/t$ at which the lattice turns insulating.
A remarkably different phenomenon can be observed in a three-band model with Hubbard-Kanamori interactions, namely the survival of the metallic phase to large interactions~\cite{deMedici2011,Isidori2019}.
We thus turn to the description this so-called Hund metal within gGut.

\subsection{Hund Metal}

The Hund coupling $J$, formally responsible for choosing the most stable atomic multiplets in molecules and solids, plays a central role in the understanding of the electronic properties in Fe-based correlated materials.
Perchance, one of the most striking phenomenologies that this interaction mediates is the existence of metallic phases that survive the onset of large Hubbard $U$ and Hund $J$ interactions~\cite{Isidori2019}.
This is the result of the careful balance between two insulating phases of different nature: a Mott insulator on the one hand, and a Hund insulator on the other.
In contrast to the Mott-to-band insulator scenario in the previous model, here both insulating phases are correlated, and can be understood by invoking the atomic limit.
In this light, the Mott insulator is a spatially homogeneous state of high-spin atoms, while the Hund insulator is a spatially-inhomogeneous charge-disproportionate mix of different atomic multiplets.

We investigate whether the gGut approximation can capture this delicate balance between correlated insulating states, resulting in an interaction-resilient metal, in the simplest example of three orbitals at a partial filling of $n = 2$ electrons per atom. Specifically, the local Hamiltonian in this case can be written compactly as
\beal
    H_{loc,i} &=-\mu\,\sum_{a=1}^3\,\sum_\sigma\, n_{ia\sigma} +(U-3J)\;\frac{\;n^2\;}{2}\\
    &\qquad -J\bigg(2 \mathbf{S}^2+\frac{1}{2}\,\mathbf{L}^2\bigg),\label{eq:3bandKanamori}
\eal
\begin{figure}
    \centering
    \includegraphics[width=0.5\textwidth]{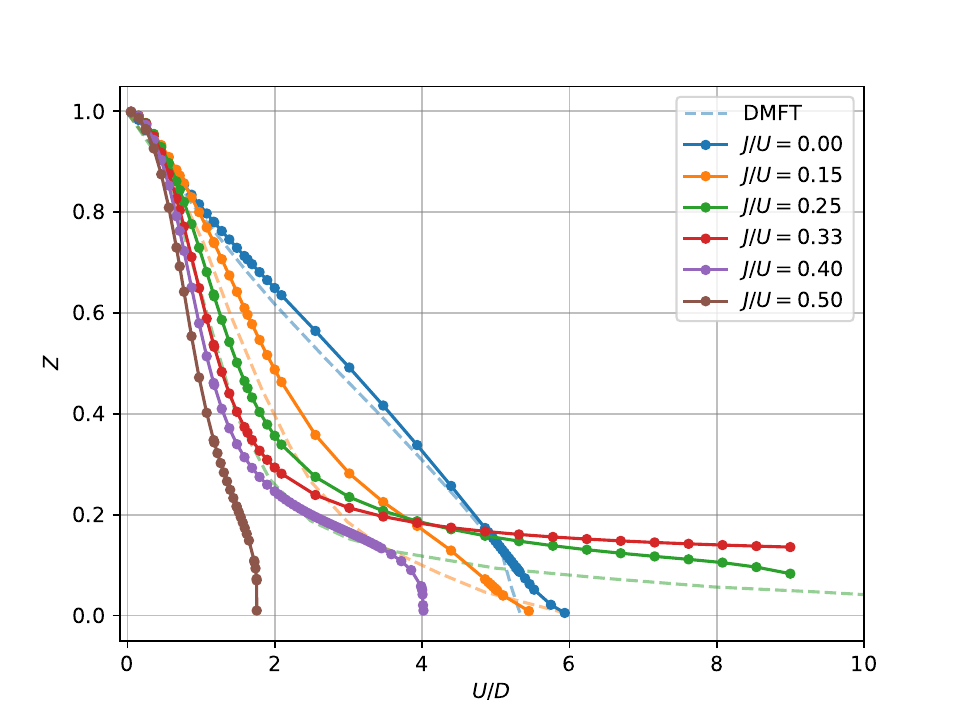}
    \caption{Quasiparticle renormalization factor $Z$ vs Hubbard repulsion $U$ for the three-band Kanamori model in Eq.~\eqref{eq:3bandKanamori}. Results are shown for two electrons per atom, for different ratios of the Hund coupling to Hubbard repulsion $J/U$. Clear differences between regular Mott insulator ($J/U < \frac{1}{3}$), Hund metal ($J/U = \frac{1}{3}$) and Hund insulator phases ($J/U > \frac{1}{3}$) can be observed. For comparison, reference results from DMFT calculations in ref.~\cite{deMedici2011} are shown as dashed lines.}
    \label{fig:janus}
\end{figure}
where $n$ is the total particle number operator, $\mathbf{S} = \frac{1}{2}\sum_{a\sigma\sigma'}c^\dagger_{a\sigma}\boldsymbol{\tau}_{\sigma\sigma'}c^\dagga_{a\sigma'}$ is the local spin operator in a given atom (with Pauli matrices $\boldsymbol{\tau}=\left\{\tau_x,\tau_y,\tau_z\right\}$). The operator $\mathbf{L} = \sum_{ab\sigma}c^\dagger_{a\sigma}\boldsymbol{\ell}_{ab}c^\dagga_{b\sigma}$ is the atomic angular momentum, where $\boldsymbol{\ell}=\left\{l_x,l_y,l_z\right\}$ with $l_\alpha$, $\alpha=x,y,z$, the angular momentum operators projected in the three-orbital basis, which we assume describe $p$ orbitals, although they could as well represent $t_{2g}$ ones. The local configurations can be labelled by the number of electrons $n$, total spin $S$ and total angular momentum $L$ and have energy $E_{loc}(n,S,L)$. The lowest states for $n=1,2,3$ have energies
\bealn
E(1,1/2,1) &= -\mu +\frac{1}{2}(U-3J) -\frac{5}{2}J\,,\\
E(2,1,1) &= -2\mu + 2(U-3J) -5J\,,\\
E(3,3/2,0) &= -3\mu +\frac{9}{2}(U-3J)- \frac{15}{2}J\,,
\eal
so that the effective Hubbard interaction projected in that subspace reads  
\beal 
U_\text{eff} &= E(1,1/2,1)+E(3,3/2,0)-2E(2,1,1)\\
&=U-3J\,.\label{U-eff}
\eal
It follows that $J>0$ effectively reduces $U$ to such an extent that, at 
$J>U/3$, the model prefers to have sites with $n=1$ and $n=3$ rather than 
the average $n=2$. Even though such extreme circumstance is maybe unrealistic in the lattice scenario, still the net effect of a $J<U/3$ is to make a metallic state survive longer than expected and with the anomalous property that each site is with highest probability in the configuration favoured by Hund's rules, contrary to the conventional metal at small $J$.  

\begin{figure*}

\subfloat[Low interaction regime ($U/D = 0.05$).\label{fig:Mults_SmallU}]{%
  \includegraphics[width=\columnwidth]{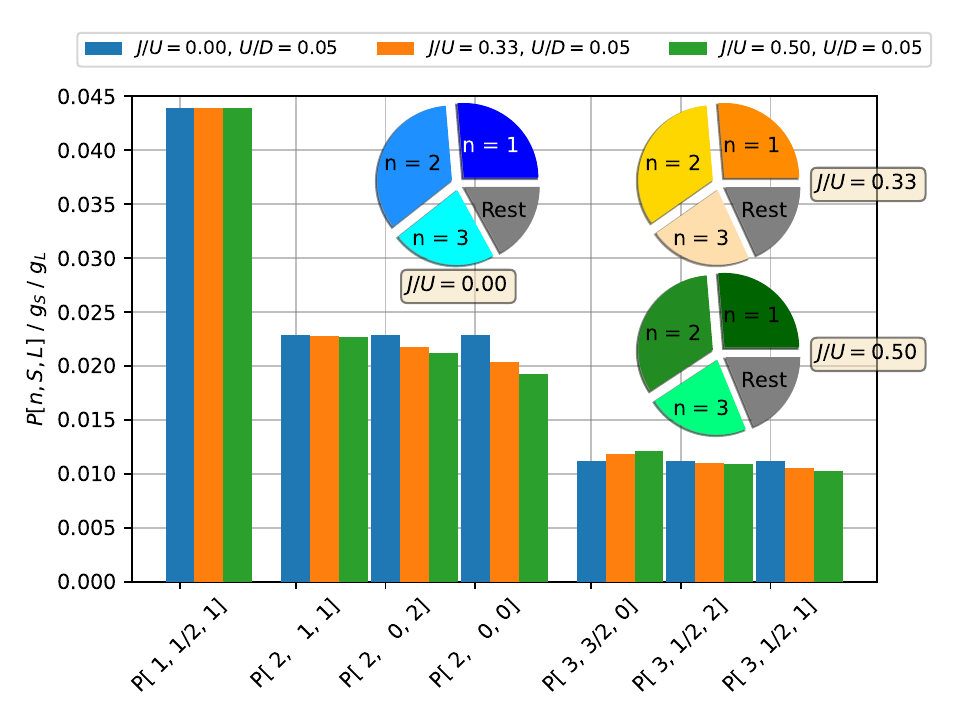}%
}\hspace*{\fill}%
\subfloat[High interaction regime.\label{fig:MultsLargeU}]{%
  \includegraphics[width=\columnwidth]{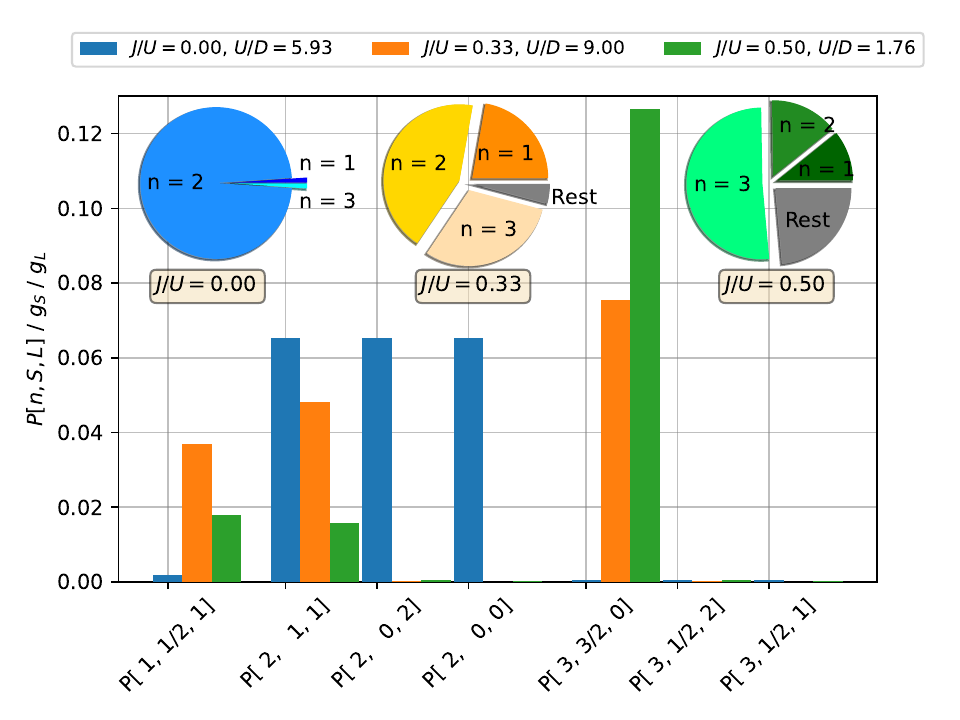}%
}
\caption{Impurity multiplet structure for the converged impurity models in the three band Hubbard-Kanamori model at one-third filling, different Hubbard interaction strengths and three Hund-to-Hubbard rations $J/U = $ 0 (blue), 0.33 (orange) and 0.5 (green). Shown are the probabilities for different impurity multiplets $\ket{n, S, L}$ defined by the impurity number of particles $n$, total spin $S$ and total orbital angular momentum $L$. These probabilites, shown in the bar plots, are weighted by the inverse degeneracies of each multiplet ($g_S = (2S+1)$, $g_L = (2L+1)$). The inset pie-charts show the probability distribution of the number of particles alone for each $J/U$ ration.} \label{fig:Multiplets}
\end{figure*}

We carry out gGut simulations on this three-band Hamiltonian at one-third filling for different ratios $J/U$, using nine baths (six ghosts). In Fig.~\ref{fig:janus} we report the quasi-particle renormalization factor of one of the equivalent bands as a function of $U/D$.
As can be seen in the figure, the gGut approximation perfectly captures the qualitative picture presented in Ref.~\cite{deMedici2011,Isidori2019}.
Starting from a regular Mott transition at $J/U = 0$, increasing the Hund's coupling ratio starts showing the typical initial $Z$ decay with subsequent metallic plateau, which eventually drops into a high angular momentum Mott insulating phase.
At large $J/U > \frac{1}{3}$ the lattice shows a similar behaviour with increasing $U/t$, where the final insulating state is instead a Hund insulator.

Besides correctly describing both correlated insulating phases, the gGut approximation also successfully captures the Hund metallic behavior.
At the precise ratio $J/U = \frac{1}{3}$, thus $U_\text{eff}=0$ in  Eq.~\eqref{U-eff}, the model exhibits a finite $Z$, i.e. a metallic phase, which essentially plateaus for a large range of interaction strengths, all the way to interactions over nine times larger than the lattice's bandwidth.
While, as was the case of the orbital-selective Mott transition, the quasiparticle renormalization factors $Z$ are systematically larger than those obtained by dynamical mean-field theory~\cite{deMedici2011}, the results in Fig.~\ref{fig:janus} clearly show that the gGut approximation can perfectly describe the Hund metal phenomenology.

We can further analyze the gGut data, better illustrating the differences between the Mott insulating, Hund insulating and Hund metallic phases, by looking at the probability distribution of the local electronic configurations for the different atomic multiplets in the $n = 1,2,3$ sectors.
This can be done by projecting the converged impurity model wave function in the gGut Ansatz into the $\ket{n,S,L}$ multiplets of interest, identified by their particle number ($n$), total spin angular momentum ($S$) and total orbital angular momentum ($L$).
For $n = 1$, the only possible multiplet is $\ket{1,1/2,1}$, for $n = 2$ the local electronic configuration can be either $\ket{2,1,1}$, $\ket{2,0,2}$ or $\ket{2,0,0}$, and finally for $n = 3$ the possible multiplets are $\ket{3, 3/2, 0}$, $\ket{3, 1/2, 2}$ and $\ket{3, 1/2, 1}$.
At one-third filling, i.e. an average local particle number of $n = 2$, the Hubbard repulsion will favor the $n = 2$ multiplets, with all different $S$ and $L$ quantum numbers being equally likely.
Meanwhile, the Hund term will select the multiplet with highest spin, namely $S = 3/2$, and hence will eventually favor a charge disproportionated configuration where half of the atoms have $n = 3$ and the other half have $n = 1$.
This different multiplet selection is another way to interpret the effective Hubbard interaction $U_{\mathrm{eff}}$ in Eq.~\eqref{U-eff}.

In Fig.~\ref{fig:Multiplets}, we show the probability distribution as bar plots for the aforementioned multiplets in the converged impurity models of the gGut solution of the three band model at different interaction regimes.
To analyze the differences between the Mott insulator, Hund insulator and Hund metal, we consider the gGut solutions at three different Hund to Mott ratios: $J/U = 0$ (pure Hubbard scenario), $J/U = 0.33$ (balanced competiton between Hubbard and Hund), and $J/U = 0.5$ (dominating Hund scenario).
For a more transparent comparison, we devide each multiplet probability by its degeneracy, which is given by $(2S+1)(2L+1)$.
Moreover, we show the pure occupation-number probability distribution (i.e. integrating out the $S$ and $L$ quantum numbers) as inset pie-charts for each $J/U$ ratio.
We consider two interaction strength regimes: a small interaction regime ($U/D = 0.05$, left panel) in which for all $J/U$ ratios the system is a weakly correlated metal, and the high interaction regime (right panel) where for $J/U \neq 0.33$ we tune the Hubbard repulsion strength to be just before the metal-insulator transition, and for $J/U = 0.33$ we ramp up the interaction to $U/D = 9$.

In the low interaction regime (left panel, $U/D = 0.05$), the multiplet distribution is quite similar for all $J/U$ ratios.
Exactly at $J/U = 0$, all the $(S,L)$ multiplets for a given $n$ are equally likely, as they should, and turning $J/U$ on does not change this significantly with these small interaction strengths.
Examining the $n$ probability distributions in the inset pie charts, the $n = 2$ sector seems to be slightly favored, as makes sense for a positive $U$ at one-third filling, but the distributions for $n = 1,2,3$ are approximately equivalent.
Occupations of $n = 0$ or $n > 3$ are comparatively disfavored through the Hubbard repulsion.
This is a consistent picture for a weakly correlated metal for all $J/U$ ratios.

In contrast to that, in the high interaction regime (right panel), the distinction between the $J/U$ ratios, and hence the outcome of the competition between the Hubbard and Hund energy scales, is more dramatic.
For $J/U = 0$, at a $U/D$ interaction shy of the Mott transition, the impurity is almost exclusively in the $n = 2$ sector (cf. the leftmost inset pie chart in the right panel), while all three corresponding $(S,L)$ multiplets are still equally likely.
This is decidedly different for the metal shy of the Hund insulator transition at $J/U = 0.5$, where the $n = 3$ is the dominant one, while $n = 1,2$ still contribute sizeably to the atomic electronic configuration (cf. the rightmost inset pie chart in the right panel).
Moreover, inside the $n=3$ sector at $J/U = 0.5$, the impurity is almost exclusively in the $S = 3/2$ multiplet, showing a perfect following of the first Hund rule.
In the regime where both Hubbard and Hund energy scales are evenly balanced, $J/U = 0.33$, the $n$ probability distribution resembles the most the weakly correlated metal (cf. the central pie chart in the right panel, compared to the pie charts in the left panel).
There are noticeable quantitative deviations, most importantly the $n = 3$ sector increasing in weight in detriment of $n = 1$ and the $n>3 ,n = 0$ contributions, but the $n = 2$ sector is still the dominant one.
This is not completely surprising, since the quasi-particle renormalization factor in Fig.~\ref{fig:janus} plateaus into a metallic value in this regime.
Notwithstanding these similarities however, the detailed multiplet distribution is fundamentally different between this Hund metallic phase and the weakly correlated metal phase in the left panel of Fig.~\ref{fig:Multiplets}.
Indeed, each $n$ sector is essentially dominated by a single $(S,L)$ multiplet, precisely the one maximizing the total spin, and fulfilling thus the first Hund rule.

The analysis of the multiplet probability distributions in Fig.~\ref{fig:Multiplets} gives thus a clear picture as to how a correlated metallic phase can emerge through the competition of two markedly distinct insulating phases, yet still showing some degree of resemblance with both of them.
Moreover, our study shows that the gGut Ansatz perfectly captures this competition and its underlying phenomenology.
This is done in excellent agreement with established approaches such as DMFT, despite the fact that gGut is formulated around the computationally and physically much simpler 1-RDM. 

\section{Conclusions and Outlook}

We have tested the reliability of the ghost Gutzwiller (gGut) approximation in a wide range of different strongly correlated phenomena occurring in multi-orbital systems, including orbital-selective Mott transitions, Mott-to-band insulator transitions as well as Hund metallicity resilient to high interactions.
Using a minimal number of additional degrees of freedom (baths/ghosts), we have shown that for a moderate computational cost, gGut can provide accurate results comparable to well-established approaches such as dynamical mean-field theory (DMFT), as well give a correct picture for the underlying physical phenomenology.
The main figure of merit concerns the fact that gGut can provide an accurate description of spectral functions despite being formulated as a self-consistency around the computationally inexpensive static one-body reduced density matrix (1-RDM). 
Consequently, this method represents an excellent approach to perform rapid yet reliable phase space explorations in model Hamiltonians, to chart out possible phases of matter which can then be investigated with more accurate methodologies.

The variational nature of the gGut approximation and its formulation around the 1-RDM make it particularly attractive for two extensions beyond the equilibrium properties of model Hamiltonians.
First, the 1-RDM is significantly easier to accurately approximate than the one-body Green's function, particularly in \emph{ab initio} calculations.
This, together with the impressively accurate Green's functions in gGut, on par with DMFT, makes the modelling of realistic materials via an LDA+gGut scheme particularly promising.
Finally, the variational formulation of gGut makes it especially amenable to generalizations to non-equilibrium, as is indeed the case with the standard Gutzwiller approximation~\cite{Schiro2010,Sandri2013,Fabrizio2013}.
Despite its significantly simplified structure when compared to non-equilibrium DMFT~\cite{Aoki2014}, here as well the Gutzwiller-based formalism can provide a reliable description of electron dynamics after quenches.
Given the significant improvement of equilibrium results when adding the ghost degrees of freedom, extending gGut to the non-equilibrium regime presents an exciting avenue of research, in which initial results are highly encouraging~\cite{GuerciThesis,Guerci2023}.
Taking everything into account, the gGut approximation promises to be a flexible and reliable tool to study correlated materials in and out of equilibrium.

\emph{Note Added}: Durint the reviewing process, two papers have come forth studying the convergence of gGut in the number of baths for the multi-orbital setting~\cite{lee2023}, and the formal connection of gGut with the density-matrix embedding theory (DMET)~\cite{lanata2023}.

\section*{Acknowledgements}
We gratefully acknowledge insightful discussions with Adriano Amaricci, Massimo Capone, Jan Skolimowski and Daniele Guerci. 

\section*{Appendix A: Fit of quasiparticle potential $\lambda$}

The first step in each gGut iteration involves the search for a quasiparticle potential $\lambda$ which, for a given renormalization $R$, defines a quasiparticle Hamiltonian $H_{qp}$ whose local 1-RDM matches the impurity 1-RDM in the previous iteration.
At its core, this problem can be formulated as an optimization, or fitting, problem for the matrix $\lambda$.
In our current implementation, we treat this as a multi-root finding problem, with cost function

\begin{equation}
    \vec{f}(\lambda) = \Delta^{I}[\lambda] - \Delta^{I-1} \overset{!}{=} \vec{0}.
    \label{eq:MultiRootCost}
\end{equation}

where $\vec{f}$ is a vectorized representation of the matrix difference in the RHS.
The solution of this multi-root problem is found iteratively using GSL's implementation of the hybrid Powel method~\cite{gsl}, which requires providing an expression for the Jacobian of the cost function with respect to the optimization parameters.
In other words, we need to evaluate the derivatives of the components of $\vec{f}$ with respect to the elements in $\lambda$, which in turn means we need the derivatives of the ground state 1-RDM with respect to the local quasiparticle potential.
These can be expressed analytically and evaluated numerically at the same cost and with the same precision than the 1-RDM itself.
For that purpose, we will suppose that $\lambda$ is parametrized by some basis of Hermitian expansion matrices $\left\{L_i\right\}$, such that we can write

\begin{equation}
    \lambda[\left\{l_i\right\}] = \sum_i l_i\ L_i,
    \label{eq:lambd_decomp}
\end{equation}

where the expansion coefficients $l_i$ will be assumed real for simplicity.
This notation will simplify expressing the $\Delta
^I$ derivatives, as we will write them in terms of the real coefficients $l_i$.
Moreover, a judicious choice of the expansion matrices $\left\{L_i\right\}$ will simplify the implementation of the symmetries of relevance in the model of interest, greatly increasing the convergence speed of the fit as well as that of the whole gGut self-consistency.

\subsubsection{Derivatives of $\Delta$}

Since the quasiparticle Hamiltonian is non-interacting, the local 1-RDM $\Delta^I$ can be written in terms of the quasiparticle potential $\lambda$ as

\begin{equation}
    \Delta^I[\left\{l_i\right\}] = \frac{1}{V_{BZ}}\sum_k\ F( R^\dagger \cdot \epsilon_k \cdot R - \lambda[\left\{l_i\right\}] ),
    \label{eq:Delta_From_Lambda}
\end{equation}

where $F$ denotes the Fermi distribution at the temperature of interest.
The above equation sets the formal functional dependence of $\Delta^I$ with respect to the expansion coefficients $l_i$.
Since the momentum sum is a linear operator, it commutes with the derivatives with respect to $l_i$, and hence we need just concern ourselves with the derivative of the Fermi distribution.
Now, the $F$ is evaluated over a matrix, hence we compute it by changing to the eigenbasis of the argument $R^\dagger \cdot \epsilon_k \cdot R- \lambda$.
Let the eigenvectors and eigenvalues be given by $\ket{v_{k,n}[\left\{l_i\right\}]}$ and $\beta_{k,n}[\left\{l_i\right\}]$, such that

\begin{equation}
    \left(R^\dagger \cdot \epsilon_k \cdot R- \lambda\right)\ket{v_{k,n}} = \beta_{k,n}\ket{v_{k,n}},
    \label{eq:DispEvals}
\end{equation}

where for clarity of notation we have omitted the explicit dependency on $\left\{l_i\right\}$.
By taking the derivative of Eq.~\eqref{eq:DispEvals} with respect to a given $l_i$, and following essentially the same arguments as when deriving Rayleigh-Schr\"odinger perturbation theory, we can find the expressions for the derivatives of the eigenvectors and eigenvalues in the absence of degeneracies.
These follow

\begin{equation}
    \begin{split}
        \frac{\partial}{\partial l_i}\beta_{k,n} &= -\braket{v_{k,n}|\frac{\partial}{\partial l_i}\lambda|v_{k,n}},\\
        \frac{\partial}{\partial l_i}\ket{v_{k,n}} &= \sum_{m\neq n} \frac{\braket{v_{k,m}|\frac{\partial}{\partial l_i}\lambda|v_{k,n}}}{\beta_{k,m}-\beta_{k,n}}\ket{v_{k,m}}.
    \end{split}
    \label{eq:DerEvEvvWithLambda}
\end{equation}

Hence, writing the 1-RDM at a given momentum vector $k$, denoted as $\Delta^I_k$, in the eigenbasis of $R^\dagger\cdot\epsilon_k\cdot R - \lambda$ as $\Delta^I_k = \sum_n F(\beta_{k,n})\ket{v_{k,n}}\bra{v_{k,n}}$, we can express its derivative as

\begin{equation}
    \frac{\partial}{\partial l_i} \Delta^I_k = \sum_{n,m} \tilde\delta^k_{n,m}\ \ket{v_{k,n}}\bra{v_{k,m}},
    \label{eq:DeltaDer}
\end{equation}

where we have defined the $\tilde\delta^k$ matrix as

\begin{equation}
    \tilde\delta^k_{n,m} = 
    \left\{\begin{array}{cc}
         -F'(\beta_{k,n}) \braket{v_{k,n}|\frac{\partial}{\partial l_i}\lambda|v_{k,n}}, & n = m \\
         \frac{F(\beta_{k,n}-F(\beta_{k,m})}{\beta_{k,m}-\beta_{k,n}}\braket{v_{k,n}|\frac{\partial}{\partial l_i}\lambda|v_{k,m}}, & n \neq m
    \end{array}\right.
    \label{eq:DeltaDerFact}
\end{equation}

In the previous equation, we evaluate the derivatives of $\lambda$ following Eq.~\eqref{eq:lambd_decomp} as $\frac{\partial}{\partial l_i}\lambda = L_i$, and $F'(E)$ is the derivative of the Fermi distribution with respect to energy.
To make the latter a numerically stable magnitude, we assign a small but finite temperature to the evaluation of these derivatives.
Finally, the derivative of the local quasiparticle 1-RDM can then be evaluated by integrating Eq.~\eqref{eq:DeltaDer} over $k$. 
In the case of a degenerate spectrum of the $k$-Hamiltonian $R^\dagger\cdot\epsilon_k\cdot R - \lambda$, we can approximate the quotient in the off-diagonal components of $\tilde\delta^k$ with the derivative of $F$.

\subsubsection{$\Delta$ parametrization}

As mentioned above, the choice of expansion matrices $L_i$ in the expansion of $\lambda$, cf. Eq.~\eqref{eq:lambd_decomp}, can significantly improve the convergence of both the fit and the overall gGut self-consistency.
For the former, it defines the directions in parameter space along which the hybrid Powell method will be defined, and for the latter it can help enforce the right system symmetries.
These symmetries are direcly imposed on $\lambda$, but have of course a direct effect on the elements of $\Delta^I$.

\begin{figure*}

\subfloat[Different band widths.\label{fig:bath_2b}]{%
  \includegraphics[width=\columnwidth]{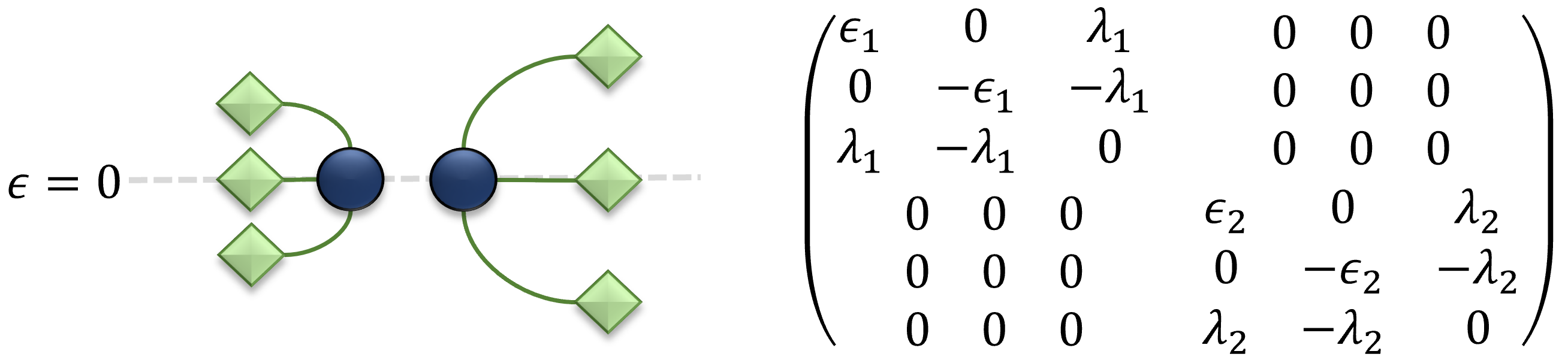}%
}\hspace*{\fill}%
\subfloat[Crystal field splitting\label{fig:bath_2b_cf}]{%
  \includegraphics[width=\columnwidth]{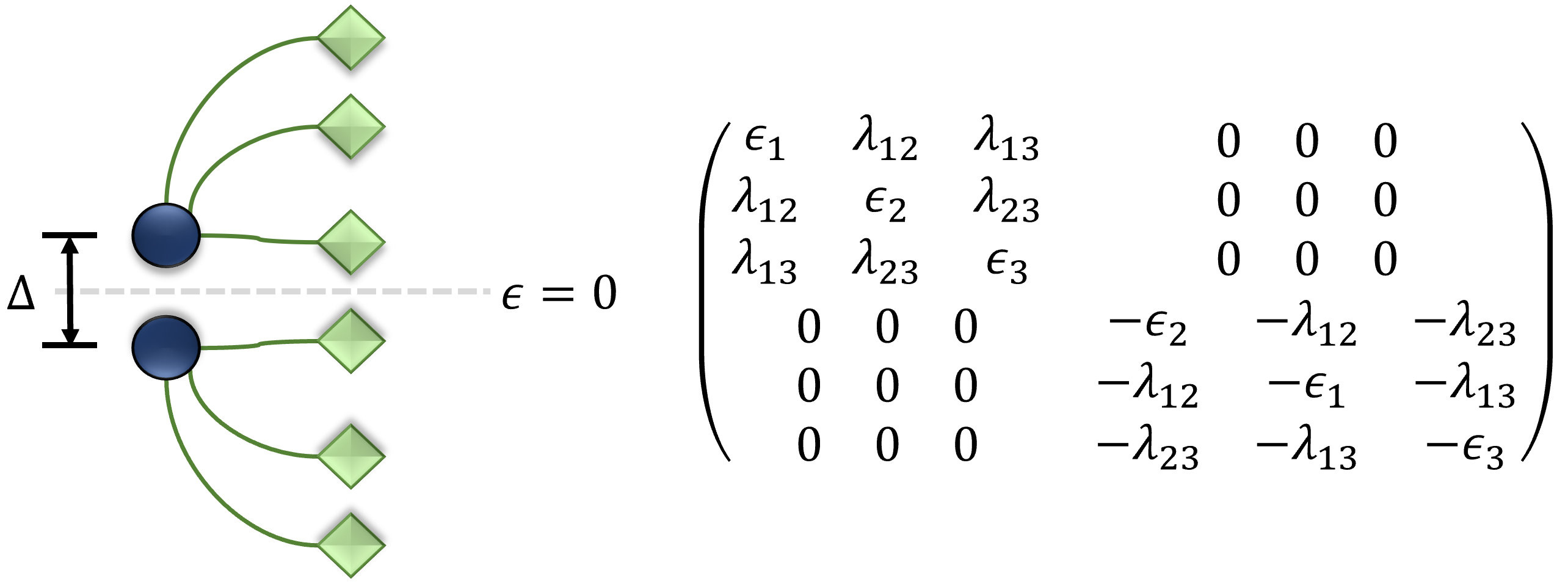}%
}
\caption{Bath parametrizations for the two-band models studied in the main paper. Left panel: two-band model at half-filling with different band widths for each band. Right panel: two-band model at half-filling with crystal field splitting.} \label{fig:bath_params}
\end{figure*}

In points of phase space where no particular symmetries are preserved, the only restriction that can be imposed on $\lambda$ is its Hermiticity.
In this work, we chose the generalized Gell-Mann matrices~\cite{Bertlmann2008} as basis in this case.
When particular symmetries arise in the system of interest, the number of expansion matrices can be further reduced.
The most common case in our current study is that of particle-hole symmetry.
Here, one can formally collect all bath orbitals in pairs, up to possibly a single orbital in case of an odd number of effective orbitals, which is paired with itself.
By particle-hole symmetry, the on-site energies of any given bath orbital is minus the energy of its pair.
Further, there can be no hopping term in $\lambda$ coupling an orbital with its particle-hole symmetric pair, and if $a$ hops to $b$ with some amplitude $\lambda_{a,b}$, then their particle-hole symmetric partners must have the same hopping amplitude between each other.

Another bath parametrization of interest pertains to the multi-orbital case.
It is not uncommon for the orbitals in the local Hamiltonian to be equivalent to each other, in which case a natural $\lambda$ parametrization involves splitting the bath orbitals among all local impurity orbitals, such that each impurity is coupled to an equivalent bath.
Further, in the absence of explicit inter-orbital hoppings in the impurity, one can further restrict these individual bath groups such that they stay decoupled from each other.

In summary, it is important to choose the right $\lambda$ parametrization to ensure a reasonable convergence of the gGut self-consistency.
The parametrizations employed for the two-band models in the main paper, both studied at half-filling and hence presenting particle-hole symmetry, are summarized and graphically represented in Fig.~\ref{fig:bath_params}.
For the three band model, since we considered the one-third filling case, there are no symmetries besides the equivalency and decoupling between each band.
Hence, the bath parametrization corresponded to $\lambda \in \begin{pmatrix} h & 0 & 0 \\ 0 & h & 0 \\ 0 & 0 & h\end{pmatrix}$, where $h$ is a general, hermitian 3$\times$3 matrix.

\section*{Appendix B: Derivative of $\sqrt{\Delta^I(\mathbb{I}-\Delta^I)}$}

For completeness, we briefly go over the evaluation of the derivatives of $\sqrt{\Delta^I(\mathbb{I}-\Delta^I)}$ that enters the computation of $\lambda^c$ in Eq.~\eqref{eq:lambd_c}.
A useful trick to perform this derivative involves defining the auxiliary matrix $X = \sqrt{\Delta^I(\mathbb{I}-\Delta^I)}$, and writing

\begin{equation}
    \begin{split}
        \mathrm{d}(X\cdot X) &= \mathrm{d}(\Delta^I(\mathbb{I}-\Delta^I)),\\
        X\cdot\mathrm{d}X + \mathrm{d}X\cdot X &= \mathrm{d}(\Delta^I(\mathbb{I}-\Delta^I)).
    \end{split}
    \label{eq:Xder}
\end{equation}

Now, the right hand side can be easily evaluated, since it does not involve the square root of a matrix.
Eq.~\eqref{eq:Xder} is an example of a Sylvester equation, which can be solved as a linear system of equations upon vectorizing the matrix $\mathrm{d}X$, which incidentally is the derivative we are after.
The vectorized equation reads

\begin{equation}
    \left(\mathbb{I}\otimes X + X^t\otimes \mathbb{I}\right)\cdot \vec{\mathrm{d}X} = \vec{b},
    \label{eq:SylvesterVectorized}
\end{equation}

where $\otimes$ denotes a Kronecker product, and $\vec{b}$ is the vectorized form of $\mathrm{d}(\Delta^I(\mathbb{I}-\Delta^I))$.


\begin{thebibliography}{90}%
\makeatletter
\providecommand \@ifxundefined [1]{%
 \@ifx{#1\undefined}
}%
\providecommand \@ifnum [1]{%
 \ifnum #1\expandafter \@firstoftwo
 \else \expandafter \@secondoftwo
 \fi
}%
\providecommand \@ifx [1]{%
 \ifx #1\expandafter \@firstoftwo
 \else \expandafter \@secondoftwo
 \fi
}%
\providecommand \natexlab [1]{#1}%
\providecommand \enquote  [1]{``#1''}%
\providecommand \bibnamefont  [1]{#1}%
\providecommand \bibfnamefont [1]{#1}%
\providecommand \citenamefont [1]{#1}%
\providecommand \href@noop [0]{\@secondoftwo}%
\providecommand \href [0]{\begingroup \@sanitize@url \@href}%
\providecommand \@href[1]{\@@startlink{#1}\@@href}%
\providecommand \@@href[1]{\endgroup#1\@@endlink}%
\providecommand \@sanitize@url [0]{\catcode `\\12\catcode `\$12\catcode
  `\&12\catcode `\#12\catcode `\^12\catcode `\_12\catcode `\%12\relax}%
\providecommand \@@startlink[1]{}%
\providecommand \@@endlink[0]{}%
\providecommand \url  [0]{\begingroup\@sanitize@url \@url }%
\providecommand \@url [1]{\endgroup\@href {#1}{\urlprefix }}%
\providecommand \urlprefix  [0]{URL }%
\providecommand \Eprint [0]{\href }%
\providecommand \doibase [0]{https://doi.org/}%
\providecommand \selectlanguage [0]{\@gobble}%
\providecommand \bibinfo  [0]{\@secondoftwo}%
\providecommand \bibfield  [0]{\@secondoftwo}%
\providecommand \translation [1]{[#1]}%
\providecommand \BibitemOpen [0]{}%
\providecommand \bibitemStop [0]{}%
\providecommand \bibitemNoStop [0]{.\EOS\space}%
\providecommand \EOS [0]{\spacefactor3000\relax}%
\providecommand \BibitemShut  [1]{\csname bibitem#1\endcsname}%
\let\auto@bib@innerbib\@empty
\bibitem [{\citenamefont {Georges}\ \emph {et~al.}(1996)\citenamefont
  {Georges}, \citenamefont {Kotliar}, \citenamefont {Krauth},\ and\
  \citenamefont {Rozenberg}}]{Kotliar1996}%
  \BibitemOpen
  \bibfield  {author} {\bibinfo {author} {\bibfnamefont {A.}~\bibnamefont
  {Georges}}, \bibinfo {author} {\bibfnamefont {G.}~\bibnamefont {Kotliar}},
  \bibinfo {author} {\bibfnamefont {W.}~\bibnamefont {Krauth}},\ and\ \bibinfo
  {author} {\bibfnamefont {M.~J.}\ \bibnamefont {Rozenberg}},\ }\bibfield
  {title} {\bibinfo {title} {Dynamical mean-field theory of strongly correlated
  fermion systems and the limit of infinite dimensions},\ }\href@noop {}
  {\bibfield  {journal} {\bibinfo  {journal} {Rev. Mod. Phys.}\ }\textbf
  {\bibinfo {volume} {68}},\ \bibinfo {pages} {13} (\bibinfo {year}
  {1996})}\BibitemShut {NoStop}%
\bibitem [{\citenamefont {Kotliar}\ \emph {et~al.}(2006)\citenamefont
  {Kotliar}, \citenamefont {Savrasov}, \citenamefont {Haule}, \citenamefont
  {Oudovenko}, \citenamefont {Parcollet},\ and\ \citenamefont
  {Marianetti}}]{Kotliar2006}%
  \BibitemOpen
  \bibfield  {author} {\bibinfo {author} {\bibfnamefont {G.}~\bibnamefont
  {Kotliar}}, \bibinfo {author} {\bibfnamefont {S.~Y.}\ \bibnamefont
  {Savrasov}}, \bibinfo {author} {\bibfnamefont {K.}~\bibnamefont {Haule}},
  \bibinfo {author} {\bibfnamefont {V.~S.}\ \bibnamefont {Oudovenko}}, \bibinfo
  {author} {\bibfnamefont {O.}~\bibnamefont {Parcollet}},\ and\ \bibinfo
  {author} {\bibfnamefont {C.}~\bibnamefont {Marianetti}},\ }\bibfield  {title}
  {\bibinfo {title} {Electronic structure calculations with dynamical
  mean-field theory},\ }\href@noop {} {\bibfield  {journal} {\bibinfo
  {journal} {Rev. Mod. Phys.}\ }\textbf {\bibinfo {volume} {78}},\ \bibinfo
  {pages} {865} (\bibinfo {year} {2006})}\BibitemShut {NoStop}%
\bibitem [{\citenamefont {Kotliar}\ and\ \citenamefont
  {Abrahams}(2001)}]{Kotliar2001b}%
  \BibitemOpen
  \bibfield  {author} {\bibinfo {author} {\bibfnamefont {G.}~\bibnamefont
  {Kotliar}}\ and\ \bibinfo {author} {\bibfnamefont {E.}~\bibnamefont
  {Abrahams}},\ }\bibfield  {title} {\bibinfo {title} {Correlated electrons in
  delta-plutonium within a dynamical mean-field picture},\ }\href@noop {}
  {\bibfield  {journal} {\bibinfo  {journal} {Nature}\ }\textbf {\bibinfo
  {volume} {410}},\ \bibinfo {pages} {793} (\bibinfo {year}
  {2001})}\BibitemShut {NoStop}%
\bibitem [{\citenamefont {Arita}\ \emph {et~al.}(2007)\citenamefont {Arita},
  \citenamefont {Held}, \citenamefont {Lukoyanov},\ and\ \citenamefont
  {Anisimov}}]{Arita2007}%
  \BibitemOpen
  \bibfield  {author} {\bibinfo {author} {\bibfnamefont {R.}~\bibnamefont
  {Arita}}, \bibinfo {author} {\bibfnamefont {K.}~\bibnamefont {Held}},
  \bibinfo {author} {\bibfnamefont {A.~V.}\ \bibnamefont {Lukoyanov}},\ and\
  \bibinfo {author} {\bibfnamefont {V.~I.}\ \bibnamefont {Anisimov}},\
  }\bibfield  {title} {\bibinfo {title} {Doped mott insulator as the origin of
  heavy-fermion behavior in ${\mathrm{liv}}_{2}{\mathrm{o}}_{4}$},\ }\href@noop
  {} {\bibfield  {journal} {\bibinfo  {journal} {Phys. Rev. Lett.}\ }\textbf
  {\bibinfo {volume} {98}},\ \bibinfo {pages} {166402} (\bibinfo {year}
  {2007})}\BibitemShut {NoStop}%
\bibitem [{\citenamefont {Shim}\ \emph {et~al.}(2007)\citenamefont {Shim},
  \citenamefont {Haule},\ and\ \citenamefont {Kotliar}}]{Shim2007}%
  \BibitemOpen
  \bibfield  {author} {\bibinfo {author} {\bibfnamefont {J.~H.}\ \bibnamefont
  {Shim}}, \bibinfo {author} {\bibfnamefont {K.}~\bibnamefont {Haule}},\ and\
  \bibinfo {author} {\bibfnamefont {G.}~\bibnamefont {Kotliar}},\ }\bibfield
  {title} {\bibinfo {title} {{Modeling the Localized-to-Itinerant Electronic
  Transition in the Heavy Fermion System CeIrIn$_5$}},\ }\href
  {https://doi.org/10.1126/science.1149064} {\bibfield  {journal} {\bibinfo
  {journal} {Science}\ }\textbf {\bibinfo {volume} {318}},\ \bibinfo {pages}
  {1615} (\bibinfo {year} {2007})}\BibitemShut {NoStop}%
\bibitem [{\citenamefont {Takizawa}\ \emph {et~al.}(2009)\citenamefont
  {Takizawa}, \citenamefont {Minohara}, \citenamefont {Kumigashira},
  \citenamefont {Toyota}, \citenamefont {Oshima}, \citenamefont {Wadati},
  \citenamefont {Yoshida}, \citenamefont {Fujimori}, \citenamefont {Lippmaa},
  \citenamefont {Kawasaki}, \citenamefont {Koinuma}, \citenamefont {Sordi},\
  and\ \citenamefont {Rozenberg}}]{Takizawa2009}%
  \BibitemOpen
  \bibfield  {author} {\bibinfo {author} {\bibfnamefont {M.}~\bibnamefont
  {Takizawa}}, \bibinfo {author} {\bibfnamefont {M.}~\bibnamefont {Minohara}},
  \bibinfo {author} {\bibfnamefont {H.}~\bibnamefont {Kumigashira}}, \bibinfo
  {author} {\bibfnamefont {D.}~\bibnamefont {Toyota}}, \bibinfo {author}
  {\bibfnamefont {M.}~\bibnamefont {Oshima}}, \bibinfo {author} {\bibfnamefont
  {H.}~\bibnamefont {Wadati}}, \bibinfo {author} {\bibfnamefont
  {T.}~\bibnamefont {Yoshida}}, \bibinfo {author} {\bibfnamefont
  {A.}~\bibnamefont {Fujimori}}, \bibinfo {author} {\bibfnamefont
  {M.}~\bibnamefont {Lippmaa}}, \bibinfo {author} {\bibfnamefont
  {M.}~\bibnamefont {Kawasaki}}, \bibinfo {author} {\bibfnamefont
  {H.}~\bibnamefont {Koinuma}}, \bibinfo {author} {\bibfnamefont
  {G.}~\bibnamefont {Sordi}},\ and\ \bibinfo {author} {\bibfnamefont
  {M.}~\bibnamefont {Rozenberg}},\ }\bibfield  {title} {\bibinfo {title}
  {Coherent and incoherent $d$ band dispersions in ${\text{srvo}}_{3}$},\
  }\href@noop {} {\bibfield  {journal} {\bibinfo  {journal} {Phys. Rev. B}\
  }\textbf {\bibinfo {volume} {80}},\ \bibinfo {pages} {235104} (\bibinfo
  {year} {2009})}\BibitemShut {NoStop}%
\bibitem [{\citenamefont {Haule}\ \emph {et~al.}(2010)\citenamefont {Haule},
  \citenamefont {Yee},\ and\ \citenamefont {Kim}}]{Haule2010}%
  \BibitemOpen
  \bibfield  {author} {\bibinfo {author} {\bibfnamefont {K.}~\bibnamefont
  {Haule}}, \bibinfo {author} {\bibfnamefont {C.-H.}\ \bibnamefont {Yee}},\
  and\ \bibinfo {author} {\bibfnamefont {K.}~\bibnamefont {Kim}},\ }\bibfield
  {title} {\bibinfo {title} {Dynamical mean-field theory within the
  full-potential methods: Electronic structure of ${\text{ceirin}}_{5}$,
  ${\text{cecoin}}_{5}$, and ${\text{cerhin}}_{5}$},\ }\href@noop {} {\bibfield
   {journal} {\bibinfo  {journal} {Phys. Rev. B}\ }\textbf {\bibinfo {volume}
  {81}},\ \bibinfo {pages} {195107} (\bibinfo {year} {2010})}\BibitemShut
  {NoStop}%
\bibitem [{\citenamefont {Park}\ \emph {et~al.}(2014)\citenamefont {Park},
  \citenamefont {Millis},\ and\ \citenamefont {Marianetti}}]{Park2014}%
  \BibitemOpen
  \bibfield  {author} {\bibinfo {author} {\bibfnamefont {H.}~\bibnamefont
  {Park}}, \bibinfo {author} {\bibfnamefont {A.~J.}\ \bibnamefont {Millis}},\
  and\ \bibinfo {author} {\bibfnamefont {C.~A.}\ \bibnamefont {Marianetti}},\
  }\bibfield  {title} {\bibinfo {title} {Computing total energies in complex
  materials using charge self-consistent dft + dmft},\ }\href@noop {}
  {\bibfield  {journal} {\bibinfo  {journal} {Phys. Rev. B}\ }\textbf {\bibinfo
  {volume} {90}},\ \bibinfo {pages} {235103} (\bibinfo {year}
  {2014})}\BibitemShut {NoStop}%
\bibitem [{\citenamefont {Haule}\ and\ \citenamefont
  {Birol}(2015)}]{Haule2015}%
  \BibitemOpen
  \bibfield  {author} {\bibinfo {author} {\bibfnamefont {K.}~\bibnamefont
  {Haule}}\ and\ \bibinfo {author} {\bibfnamefont {T.}~\bibnamefont {Birol}},\
  }\bibfield  {title} {\bibinfo {title} {Free energy from stationary
  implementation of the $\mathrm{DFT}+\mathrm{DMFT}$ functional},\ }\href@noop
  {} {\bibfield  {journal} {\bibinfo  {journal} {Phys. Rev. Lett.}\ }\textbf
  {\bibinfo {volume} {115}},\ \bibinfo {pages} {256402} (\bibinfo {year}
  {2015})}\BibitemShut {NoStop}%
\bibitem [{\citenamefont {Paul}\ and\ \citenamefont {Birol}(2019)}]{Paul2019}%
  \BibitemOpen
  \bibfield  {author} {\bibinfo {author} {\bibfnamefont {A.}~\bibnamefont
  {Paul}}\ and\ \bibinfo {author} {\bibfnamefont {T.}~\bibnamefont {Birol}},\
  }\bibfield  {title} {\bibinfo {title} {Applications of dft+ dmft in materials
  science},\ }\href@noop {} {\bibfield  {journal} {\bibinfo  {journal} {Annual
  Review of Materials Research}\ }\textbf {\bibinfo {volume} {49}},\ \bibinfo
  {pages} {31} (\bibinfo {year} {2019})}\BibitemShut {NoStop}%
\bibitem [{\citenamefont {de' Medici}\ and\ \citenamefont
  {Capone}(2017)}]{deMedici2017}%
  \BibitemOpen
  \bibfield  {author} {\bibinfo {author} {\bibfnamefont {L.}~\bibnamefont {de'
  Medici}}\ and\ \bibinfo {author} {\bibfnamefont {M.}~\bibnamefont {Capone}},\
  }\bibinfo {title} {Modeling many-body physics with slave-spin mean-field:
  Mott and hund's physics in fe-superconductors},\ in\ \href
  {https://doi.org/10.1007/978-3-319-56117-2_4} {\emph {\bibinfo {booktitle}
  {The Iron Pnictide Superconductors: An Introduction and Overview}}},\
  \bibinfo {editor} {edited by\ \bibinfo {editor} {\bibfnamefont
  {F.}~\bibnamefont {Mancini}}\ and\ \bibinfo {editor} {\bibfnamefont
  {R.}~\bibnamefont {Citro}}}\ (\bibinfo  {publisher} {Springer International
  Publishing},\ \bibinfo {address} {Cham},\ \bibinfo {year} {2017})\ pp.\
  \bibinfo {pages} {115--185}\BibitemShut {NoStop}%
\bibitem [{\citenamefont {Lanat\`a}\ \emph {et~al.}(2017)\citenamefont
  {Lanat\`a}, \citenamefont {Lee}, \citenamefont {Yao},\ and\ \citenamefont
  {Dobrosavljevi\ifmmode~\acute{c}\else \'{c}\fi{}}}]{Lanata2017}%
  \BibitemOpen
  \bibfield  {author} {\bibinfo {author} {\bibfnamefont {N.}~\bibnamefont
  {Lanat\`a}}, \bibinfo {author} {\bibfnamefont {T.-H.}\ \bibnamefont {Lee}},
  \bibinfo {author} {\bibfnamefont {Y.-X.}\ \bibnamefont {Yao}},\ and\ \bibinfo
  {author} {\bibfnamefont {V.}~\bibnamefont
  {Dobrosavljevi\ifmmode~\acute{c}\else \'{c}\fi{}}},\ }\bibfield  {title}
  {\bibinfo {title} {Emergent bloch excitations in mott matter},\ }\href
  {https://doi.org/10.1103/PhysRevB.96.195126} {\bibfield  {journal} {\bibinfo
  {journal} {Phys. Rev. B}\ }\textbf {\bibinfo {volume} {96}},\ \bibinfo
  {pages} {195126} (\bibinfo {year} {2017})}\BibitemShut {NoStop}%
\bibitem [{\citenamefont {Gutzwiller}(1963)}]{Gutzwiller1963}%
  \BibitemOpen
  \bibfield  {author} {\bibinfo {author} {\bibfnamefont {M.~C.}\ \bibnamefont
  {Gutzwiller}},\ }\bibfield  {title} {\bibinfo {title} {Effect of correlation
  on the ferromagnetism of transition metals},\ }\href
  {https://doi.org/10.1103/PhysRevLett.10.159} {\bibfield  {journal} {\bibinfo
  {journal} {Phys. Rev. Lett.}\ }\textbf {\bibinfo {volume} {10}},\ \bibinfo
  {pages} {159} (\bibinfo {year} {1963})}\BibitemShut {NoStop}%
\bibitem [{\citenamefont {Gutzwiller}(1965)}]{Gutzwiller1965}%
  \BibitemOpen
  \bibfield  {author} {\bibinfo {author} {\bibfnamefont {M.~C.}\ \bibnamefont
  {Gutzwiller}},\ }\bibfield  {title} {\bibinfo {title} {Correlation of
  electrons in a narrow $s$ band},\ }\href
  {https://doi.org/10.1103/PhysRev.137.A1726} {\bibfield  {journal} {\bibinfo
  {journal} {Phys. Rev.}\ }\textbf {\bibinfo {volume} {137}},\ \bibinfo {pages}
  {A1726} (\bibinfo {year} {1965})}\BibitemShut {NoStop}%
\bibitem [{\citenamefont {Guerci}\ \emph {et~al.}(2019)\citenamefont {Guerci},
  \citenamefont {Capone},\ and\ \citenamefont {Fabrizio}}]{guerci2019}%
  \BibitemOpen
  \bibfield  {author} {\bibinfo {author} {\bibfnamefont {D.}~\bibnamefont
  {Guerci}}, \bibinfo {author} {\bibfnamefont {M.}~\bibnamefont {Capone}},\
  and\ \bibinfo {author} {\bibfnamefont {M.}~\bibnamefont {Fabrizio}},\
  }\bibfield  {title} {\bibinfo {title} {Exciton mott transition revisited},\
  }\href@noop {} {\bibfield  {journal} {\bibinfo  {journal} {Physical Review
  Materials}\ }\textbf {\bibinfo {volume} {3}},\ \bibinfo {pages} {054605}
  (\bibinfo {year} {2019})}\BibitemShut {NoStop}%
\bibitem [{\citenamefont {Frank}\ \emph {et~al.}(2021)\citenamefont {Frank},
  \citenamefont {Lee}, \citenamefont {Bhattacharyya}, \citenamefont {Tsang},
  \citenamefont {Quito}, \citenamefont {Dobrosavljevi{\'c}}, \citenamefont
  {Christiansen},\ and\ \citenamefont {Lanat{\`a}}}]{frank2021}%
  \BibitemOpen
  \bibfield  {author} {\bibinfo {author} {\bibfnamefont {M.~S.}\ \bibnamefont
  {Frank}}, \bibinfo {author} {\bibfnamefont {T.-H.}\ \bibnamefont {Lee}},
  \bibinfo {author} {\bibfnamefont {G.}~\bibnamefont {Bhattacharyya}}, \bibinfo
  {author} {\bibfnamefont {P.~K.~H.}\ \bibnamefont {Tsang}}, \bibinfo {author}
  {\bibfnamefont {V.~L.}\ \bibnamefont {Quito}}, \bibinfo {author}
  {\bibfnamefont {V.}~\bibnamefont {Dobrosavljevi{\'c}}}, \bibinfo {author}
  {\bibfnamefont {O.}~\bibnamefont {Christiansen}},\ and\ \bibinfo {author}
  {\bibfnamefont {N.}~\bibnamefont {Lanat{\`a}}},\ }\bibfield  {title}
  {\bibinfo {title} {Quantum embedding description of the anderson lattice
  model with the ghost gutzwiller approximation},\ }\href@noop {} {\bibfield
  {journal} {\bibinfo  {journal} {Physical Review B}\ }\textbf {\bibinfo
  {volume} {104}},\ \bibinfo {pages} {L081103} (\bibinfo {year}
  {2021})}\BibitemShut {NoStop}%
\bibitem [{\citenamefont {Lee}\ \emph {et~al.}(2022)\citenamefont {Lee},
  \citenamefont {Lanat{\`a}},\ and\ \citenamefont {Kotliar}}]{lee2022}%
  \BibitemOpen
  \bibfield  {author} {\bibinfo {author} {\bibfnamefont {T.-H.}\ \bibnamefont
  {Lee}}, \bibinfo {author} {\bibfnamefont {N.}~\bibnamefont {Lanat{\`a}}},\
  and\ \bibinfo {author} {\bibfnamefont {G.}~\bibnamefont {Kotliar}},\
  }\bibfield  {title} {\bibinfo {title} {Accuracy of
  ghost-rotationally-invariant slave-boson and dynamical mean field theory as a
  function of the impurity-model bath size},\ }\href@noop {} {\bibfield
  {journal} {\bibinfo  {journal} {arXiv preprint arXiv:2212.07515}\ } (\bibinfo
  {year} {2022})}\BibitemShut {NoStop}%
\bibitem [{\citenamefont {B\"unemann}\ \emph {et~al.}(1998)\citenamefont
  {B\"unemann}, \citenamefont {Weber},\ and\ \citenamefont
  {Gebhard}}]{Bunemann1998}%
  \BibitemOpen
  \bibfield  {author} {\bibinfo {author} {\bibfnamefont {J.}~\bibnamefont
  {B\"unemann}}, \bibinfo {author} {\bibfnamefont {W.}~\bibnamefont {Weber}},\
  and\ \bibinfo {author} {\bibfnamefont {F.}~\bibnamefont {Gebhard}},\
  }\bibfield  {title} {\bibinfo {title} {Multiband gutzwiller wave functions
  for general on-site interactions},\ }\href@noop {} {\bibfield  {journal}
  {\bibinfo  {journal} {Phys. Rev. B}\ }\textbf {\bibinfo {volume} {57}},\
  \bibinfo {pages} {6896} (\bibinfo {year} {1998})}\BibitemShut {NoStop}%
\bibitem [{\citenamefont {Lanat{\`a}}\ \emph {et~al.}(2015)\citenamefont
  {Lanat{\`a}}, \citenamefont {Yao}, \citenamefont {Wang}, \citenamefont {Ho},\
  and\ \citenamefont {Kotliar}}]{lanata2015}%
  \BibitemOpen
  \bibfield  {author} {\bibinfo {author} {\bibfnamefont {N.}~\bibnamefont
  {Lanat{\`a}}}, \bibinfo {author} {\bibfnamefont {Y.}~\bibnamefont {Yao}},
  \bibinfo {author} {\bibfnamefont {C.-Z.}\ \bibnamefont {Wang}}, \bibinfo
  {author} {\bibfnamefont {K.-M.}\ \bibnamefont {Ho}},\ and\ \bibinfo {author}
  {\bibfnamefont {G.}~\bibnamefont {Kotliar}},\ }\bibfield  {title} {\bibinfo
  {title} {Phase diagram and electronic structure of praseodymium and
  plutonium},\ }\href@noop {} {\bibfield  {journal} {\bibinfo  {journal}
  {Physical Review X}\ }\textbf {\bibinfo {volume} {5}},\ \bibinfo {pages}
  {011008} (\bibinfo {year} {2015})}\BibitemShut {NoStop}%
\bibitem [{\citenamefont {Fabrizio}(2017)}]{fabrizio2017}%
  \BibitemOpen
  \bibfield  {author} {\bibinfo {author} {\bibfnamefont {M.}~\bibnamefont
  {Fabrizio}},\ }\bibfield  {title} {\bibinfo {title} {Quantum fluctuations
  beyond the gutzwiller approximation},\ }\href@noop {} {\bibfield  {journal}
  {\bibinfo  {journal} {Physical Review B}\ }\textbf {\bibinfo {volume} {95}},\
  \bibinfo {pages} {075156} (\bibinfo {year} {2017})}\BibitemShut {NoStop}%
\bibitem [{\citenamefont {Yokoyama}\ and\ \citenamefont
  {Shiba}(1987)}]{Yokoyama1987}%
  \BibitemOpen
  \bibfield  {author} {\bibinfo {author} {\bibfnamefont {H.}~\bibnamefont
  {Yokoyama}}\ and\ \bibinfo {author} {\bibfnamefont {H.}~\bibnamefont
  {Shiba}},\ }\bibfield  {title} {\bibinfo {title} {Variational monte-carlo
  studies of hubbard model. i},\ }\href {https://doi.org/10.1143/JPSJ.56.1490}
  {\bibfield  {journal} {\bibinfo  {journal} {Journal of the Physical Society
  of Japan}\ }\textbf {\bibinfo {volume} {56}},\ \bibinfo {pages} {1490}
  (\bibinfo {year} {1987})}\BibitemShut {NoStop}%
\bibitem [{\citenamefont {Capello}\ \emph {et~al.}(2005)\citenamefont
  {Capello}, \citenamefont {Becca}, \citenamefont {Fabrizio}, \citenamefont
  {Sorella},\ and\ \citenamefont {Tosatti}}]{Capello2005}%
  \BibitemOpen
  \bibfield  {author} {\bibinfo {author} {\bibfnamefont {M.}~\bibnamefont
  {Capello}}, \bibinfo {author} {\bibfnamefont {F.}~\bibnamefont {Becca}},
  \bibinfo {author} {\bibfnamefont {M.}~\bibnamefont {Fabrizio}}, \bibinfo
  {author} {\bibfnamefont {S.}~\bibnamefont {Sorella}},\ and\ \bibinfo {author}
  {\bibfnamefont {E.}~\bibnamefont {Tosatti}},\ }\bibfield  {title} {\bibinfo
  {title} {Variational description of mott insulators},\ }\href
  {https://doi.org/10.1103/PhysRevLett.94.026406} {\bibfield  {journal}
  {\bibinfo  {journal} {Phys. Rev. Lett.}\ }\textbf {\bibinfo {volume} {94}},\
  \bibinfo {pages} {026406} (\bibinfo {year} {2005})}\BibitemShut {NoStop}%
\bibitem [{\citenamefont {Lanat\`a}\ \emph {et~al.}(2012)\citenamefont
  {Lanat\`a}, \citenamefont {Strand}, \citenamefont {Dai},\ and\ \citenamefont
  {Hellsing}}]{Lanata2012}%
  \BibitemOpen
  \bibfield  {author} {\bibinfo {author} {\bibfnamefont {N.}~\bibnamefont
  {Lanat\`a}}, \bibinfo {author} {\bibfnamefont {H.~U.~R.}\ \bibnamefont
  {Strand}}, \bibinfo {author} {\bibfnamefont {X.}~\bibnamefont {Dai}},\ and\
  \bibinfo {author} {\bibfnamefont {B.}~\bibnamefont {Hellsing}},\ }\bibfield
  {title} {\bibinfo {title} {Efficient implementation of the gutzwiller
  variational method},\ }\href {https://doi.org/10.1103/PhysRevB.85.035133}
  {\bibfield  {journal} {\bibinfo  {journal} {Phys. Rev. B}\ }\textbf {\bibinfo
  {volume} {85}},\ \bibinfo {pages} {035133} (\bibinfo {year}
  {2012})}\BibitemShut {NoStop}%
\bibitem [{\citenamefont {Deng}\ \emph {et~al.}(2008)\citenamefont {Deng},
  \citenamefont {Dai},\ and\ \citenamefont {Fang}}]{Deng2008}%
  \BibitemOpen
  \bibfield  {author} {\bibinfo {author} {\bibfnamefont {X.}~\bibnamefont
  {Deng}}, \bibinfo {author} {\bibfnamefont {X.}~\bibnamefont {Dai}},\ and\
  \bibinfo {author} {\bibfnamefont {Z.}~\bibnamefont {Fang}},\ }\bibfield
  {title} {\bibinfo {title} {{LDA} + {G}utzwiller method for correlated
  electron systems},\ }\href@noop {} {\bibfield  {journal} {\bibinfo  {journal}
  {Europhysics Letters}\ }\textbf {\bibinfo {volume} {83}},\ \bibinfo {pages}
  {37008} (\bibinfo {year} {2008})}\BibitemShut {NoStop}%
\bibitem [{\citenamefont {Wang}\ \emph {et~al.}(2010)\citenamefont {Wang},
  \citenamefont {Qian}, \citenamefont {Xu}, \citenamefont {Dai},\ and\
  \citenamefont {Fang}}]{Wang2010}%
  \BibitemOpen
  \bibfield  {author} {\bibinfo {author} {\bibfnamefont {G.}~\bibnamefont
  {Wang}}, \bibinfo {author} {\bibfnamefont {Y.}~\bibnamefont {Qian}}, \bibinfo
  {author} {\bibfnamefont {G.}~\bibnamefont {Xu}}, \bibinfo {author}
  {\bibfnamefont {X.}~\bibnamefont {Dai}},\ and\ \bibinfo {author}
  {\bibfnamefont {Z.}~\bibnamefont {Fang}},\ }\bibfield  {title} {\bibinfo
  {title} {Gutzwiller density functional studies of feas-based superconductors:
  Structure optimization and evidence for a three-dimensional fermi surface},\
  }\href {https://doi.org/10.1103/PhysRevLett.104.047002} {\bibfield  {journal}
  {\bibinfo  {journal} {Phys. Rev. Lett.}\ }\textbf {\bibinfo {volume} {104}},\
  \bibinfo {pages} {047002} (\bibinfo {year} {2010})}\BibitemShut {NoStop}%
\bibitem [{\citenamefont {Yao}\ \emph {et~al.}(2012)\citenamefont {Yao},
  \citenamefont {Wang},\ and\ \citenamefont {Ho}}]{Yao2012}%
  \BibitemOpen
  \bibfield  {author} {\bibinfo {author} {\bibfnamefont {Y.~X.}\ \bibnamefont
  {Yao}}, \bibinfo {author} {\bibfnamefont {C.~Z.}\ \bibnamefont {Wang}},\ and\
  \bibinfo {author} {\bibfnamefont {K.~M.}\ \bibnamefont {Ho}},\ }\bibfield
  {title} {\bibinfo {title} {The benchmark of gutzwiller density functional
  theory in hydrogen systems},\ }\href@noop {} {\bibfield  {journal} {\bibinfo
  {journal} {International Journal of Quantum Chemistry}\ }\textbf {\bibinfo
  {volume} {112}},\ \bibinfo {pages} {240} (\bibinfo {year}
  {2012})}\BibitemShut {NoStop}%
\bibitem [{\citenamefont {Lanat\`a}\ \emph {et~al.}(2013)\citenamefont
  {Lanat\`a}, \citenamefont {Yao}, \citenamefont {Wang}, \citenamefont {Ho},
  \citenamefont {Schmalian}, \citenamefont {Haule},\ and\ \citenamefont
  {Kotliar}}]{Lanata2013}%
  \BibitemOpen
  \bibfield  {author} {\bibinfo {author} {\bibfnamefont {N.}~\bibnamefont
  {Lanat\`a}}, \bibinfo {author} {\bibfnamefont {Y.-X.}\ \bibnamefont {Yao}},
  \bibinfo {author} {\bibfnamefont {C.-Z.}\ \bibnamefont {Wang}}, \bibinfo
  {author} {\bibfnamefont {K.-M.}\ \bibnamefont {Ho}}, \bibinfo {author}
  {\bibfnamefont {J.}~\bibnamefont {Schmalian}}, \bibinfo {author}
  {\bibfnamefont {K.}~\bibnamefont {Haule}},\ and\ \bibinfo {author}
  {\bibfnamefont {G.}~\bibnamefont {Kotliar}},\ }\bibfield  {title} {\bibinfo
  {title}
  {$\ensuremath{\gamma}\mathrm{\text{\ensuremath{-}}}\ensuremath{\alpha}$
  isostructural transition in cerium},\ }\href
  {https://doi.org/10.1103/PhysRevLett.111.196801} {\bibfield  {journal}
  {\bibinfo  {journal} {Phys. Rev. Lett.}\ }\textbf {\bibinfo {volume} {111}},\
  \bibinfo {pages} {196801} (\bibinfo {year} {2013})}\BibitemShut {NoStop}%
\bibitem [{\citenamefont {Lanat\`a}\ \emph {et~al.}(2014)\citenamefont
  {Lanat\`a}, \citenamefont {Yao}, \citenamefont {Wang}, \citenamefont {Ho},\
  and\ \citenamefont {Kotliar}}]{Lanata2014}%
  \BibitemOpen
  \bibfield  {author} {\bibinfo {author} {\bibfnamefont {N.}~\bibnamefont
  {Lanat\`a}}, \bibinfo {author} {\bibfnamefont {Y.-X.}\ \bibnamefont {Yao}},
  \bibinfo {author} {\bibfnamefont {C.-Z.}\ \bibnamefont {Wang}}, \bibinfo
  {author} {\bibfnamefont {K.-M.}\ \bibnamefont {Ho}},\ and\ \bibinfo {author}
  {\bibfnamefont {G.}~\bibnamefont {Kotliar}},\ }\bibfield  {title} {\bibinfo
  {title} {Interplay of spin-orbit and entropic effects in cerium},\ }\href
  {https://doi.org/10.1103/PhysRevB.90.161104} {\bibfield  {journal} {\bibinfo
  {journal} {Phys. Rev. B}\ }\textbf {\bibinfo {volume} {90}},\ \bibinfo
  {pages} {161104} (\bibinfo {year} {2014})}\BibitemShut {NoStop}%
\bibitem [{\citenamefont {Borghi}\ \emph {et~al.}(2014)\citenamefont {Borghi},
  \citenamefont {Fabrizio},\ and\ \citenamefont {Tosatti}}]{Borghi2014}%
  \BibitemOpen
  \bibfield  {author} {\bibinfo {author} {\bibfnamefont {G.}~\bibnamefont
  {Borghi}}, \bibinfo {author} {\bibfnamefont {M.}~\bibnamefont {Fabrizio}},\
  and\ \bibinfo {author} {\bibfnamefont {E.}~\bibnamefont {Tosatti}},\
  }\bibfield  {title} {\bibinfo {title} {Gutzwiller electronic structure
  calculations applied to transition metals: Kinetic energy gain with
  ferromagnetic order in bcc fe},\ }\href
  {https://doi.org/10.1103/PhysRevB.90.125102} {\bibfield  {journal} {\bibinfo
  {journal} {Phys. Rev. B}\ }\textbf {\bibinfo {volume} {90}},\ \bibinfo
  {pages} {125102} (\bibinfo {year} {2014})}\BibitemShut {NoStop}%
\bibitem [{\citenamefont {Tian}\ \emph {et~al.}(2015)\citenamefont {Tian},
  \citenamefont {Song}, \citenamefont {Liu}, \citenamefont {Wang},
  \citenamefont {Fang},\ and\ \citenamefont {Dai}}]{Tian2015}%
  \BibitemOpen
  \bibfield  {author} {\bibinfo {author} {\bibfnamefont {M.-F.}\ \bibnamefont
  {Tian}}, \bibinfo {author} {\bibfnamefont {H.-F.}\ \bibnamefont {Song}},
  \bibinfo {author} {\bibfnamefont {H.-F.}\ \bibnamefont {Liu}}, \bibinfo
  {author} {\bibfnamefont {C.}~\bibnamefont {Wang}}, \bibinfo {author}
  {\bibfnamefont {Z.}~\bibnamefont {Fang}},\ and\ \bibinfo {author}
  {\bibfnamefont {X.}~\bibnamefont {Dai}},\ }\bibfield  {title} {\bibinfo
  {title} {Thermodynamics of the
  $\ensuremath{\alpha}\text{\ensuremath{-}}\ensuremath{\gamma}$ transition in
  cerium studied by an lda + gutzwiller method},\ }\href
  {https://doi.org/10.1103/PhysRevB.91.125148} {\bibfield  {journal} {\bibinfo
  {journal} {Phys. Rev. B}\ }\textbf {\bibinfo {volume} {91}},\ \bibinfo
  {pages} {125148} (\bibinfo {year} {2015})}\BibitemShut {NoStop}%
\bibitem [{\citenamefont {Peng}\ \emph {et~al.}(2021)\citenamefont {Peng},
  \citenamefont {Weng},\ and\ \citenamefont {Dai}}]{Peng2021}%
  \BibitemOpen
  \bibfield  {author} {\bibinfo {author} {\bibfnamefont {S.}~\bibnamefont
  {Peng}}, \bibinfo {author} {\bibfnamefont {H.}~\bibnamefont {Weng}},\ and\
  \bibinfo {author} {\bibfnamefont {X.}~\bibnamefont {Dai}},\ }\href
  {https://doi.org/10.48550/ARXIV.2111.09166} {\bibinfo {title} {Rtgw2020: A
  powerful implementation of dft + gutzwiller method}} (\bibinfo {year}
  {2021})\BibitemShut {NoStop}%
\bibitem [{\citenamefont {Ye}\ \emph {et~al.}(2022)\citenamefont {Ye},
  \citenamefont {Fang}, \citenamefont {Zhang}, \citenamefont {Zhang},
  \citenamefont {Wu}, \citenamefont {Lu}, \citenamefont {Yao}, \citenamefont
  {Wang},\ and\ \citenamefont {Ho}}]{Ye2022}%
  \BibitemOpen
  \bibfield  {author} {\bibinfo {author} {\bibfnamefont {Z.}~\bibnamefont
  {Ye}}, \bibinfo {author} {\bibfnamefont {Y.}~\bibnamefont {Fang}}, \bibinfo
  {author} {\bibfnamefont {H.}~\bibnamefont {Zhang}}, \bibinfo {author}
  {\bibfnamefont {F.}~\bibnamefont {Zhang}}, \bibinfo {author} {\bibfnamefont
  {S.}~\bibnamefont {Wu}}, \bibinfo {author} {\bibfnamefont {W.-C.}\
  \bibnamefont {Lu}}, \bibinfo {author} {\bibfnamefont {Y.-X.}\ \bibnamefont
  {Yao}}, \bibinfo {author} {\bibfnamefont {C.-Z.}\ \bibnamefont {Wang}},\ and\
  \bibinfo {author} {\bibfnamefont {K.-M.}\ \bibnamefont {Ho}},\ }\bibfield
  {title} {\bibinfo {title} {The gutzwiller conjugate gradient minimization
  method for correlated electron systems},\ }\href
  {https://doi.org/10.1088/1361-648X/ac5e03} {\bibfield  {journal} {\bibinfo
  {journal} {Journal of Physics: Condensed Matter}\ }\textbf {\bibinfo {volume}
  {34}},\ \bibinfo {pages} {243001} (\bibinfo {year} {2022})}\BibitemShut
  {NoStop}%
\bibitem [{\citenamefont {Brinkman}\ and\ \citenamefont
  {Rice}(1970)}]{Brinkman1970}%
  \BibitemOpen
  \bibfield  {author} {\bibinfo {author} {\bibfnamefont {W.~F.}\ \bibnamefont
  {Brinkman}}\ and\ \bibinfo {author} {\bibfnamefont {T.~M.}\ \bibnamefont
  {Rice}},\ }\bibfield  {title} {\bibinfo {title} {Application of gutzwiller's
  variational method to the metal-insulator transition},\ }\href
  {https://doi.org/10.1103/PhysRevB.2.4302} {\bibfield  {journal} {\bibinfo
  {journal} {Phys. Rev. B}\ }\textbf {\bibinfo {volume} {2}},\ \bibinfo {pages}
  {4302} (\bibinfo {year} {1970})}\BibitemShut {NoStop}%
\bibitem [{\citenamefont {Fulde}(1995)}]{Fulde1995_gutz}%
  \BibitemOpen
  \bibfield  {author} {\bibinfo {author} {\bibfnamefont {P.}~\bibnamefont
  {Fulde}},\ }\bibinfo {title} {Correlations in atoms and molecules},\ in\
  \href {https://doi.org/10.1007/978-3-642-57809-0_8} {\emph {\bibinfo
  {booktitle} {Electron Correlations in Molecules and Solids}}}\ (\bibinfo
  {publisher} {Springer Berlin Heidelberg},\ \bibinfo {address} {Berlin,
  Heidelberg},\ \bibinfo {year} {1995})\ pp.\ \bibinfo {pages}
  {151--188}\BibitemShut {NoStop}%
\bibitem [{\citenamefont {Caffarel}\ and\ \citenamefont
  {Krauth}(1994)}]{Caffarel1994}%
  \BibitemOpen
  \bibfield  {author} {\bibinfo {author} {\bibfnamefont {M.}~\bibnamefont
  {Caffarel}}\ and\ \bibinfo {author} {\bibfnamefont {W.}~\bibnamefont
  {Krauth}},\ }\bibfield  {title} {\bibinfo {title} {Exact diagonalization
  approach to correlated fermions in infinite dimensions: Mott transition and
  superconductivity},\ }\href@noop {} {\bibfield  {journal} {\bibinfo
  {journal} {Phys. Rev. Lett.}\ }\textbf {\bibinfo {volume} {72}},\ \bibinfo
  {pages} {1545} (\bibinfo {year} {1994})}\BibitemShut {NoStop}%
\bibitem [{\citenamefont {Koch}\ \emph {et~al.}(2008)\citenamefont {Koch},
  \citenamefont {Sangiovanni},\ and\ \citenamefont {Gunnarsson}}]{Koch2008}%
  \BibitemOpen
  \bibfield  {author} {\bibinfo {author} {\bibfnamefont {E.}~\bibnamefont
  {Koch}}, \bibinfo {author} {\bibfnamefont {G.}~\bibnamefont {Sangiovanni}},\
  and\ \bibinfo {author} {\bibfnamefont {O.}~\bibnamefont {Gunnarsson}},\
  }\bibfield  {title} {\bibinfo {title} {Sum rules and bath parametrization for
  quantum cluster theories},\ }\href@noop {} {\bibfield  {journal} {\bibinfo
  {journal} {Phys. Rev. B}\ }\textbf {\bibinfo {volume} {78}},\ \bibinfo
  {pages} {115102} (\bibinfo {year} {2008})}\BibitemShut {NoStop}%
\bibitem [{\citenamefont {Liebsch}\ and\ \citenamefont
  {Ishida}(2011)}]{Liebsch2011}%
  \BibitemOpen
  \bibfield  {author} {\bibinfo {author} {\bibfnamefont {A.}~\bibnamefont
  {Liebsch}}\ and\ \bibinfo {author} {\bibfnamefont {H.}~\bibnamefont
  {Ishida}},\ }\bibfield  {title} {\bibinfo {title} {Temperature and bath size
  in exact diagonalization dynamical mean field theory},\ }\href@noop {}
  {\bibfield  {journal} {\bibinfo  {journal} {J. Phys.-Condens. Mat.}\ }\textbf
  {\bibinfo {volume} {24}},\ \bibinfo {pages} {053201} (\bibinfo {year}
  {2011})}\BibitemShut {NoStop}%
\bibitem [{\citenamefont {Knizia}\ and\ \citenamefont
  {Chan}(2012)}]{Knizia2012}%
  \BibitemOpen
  \bibfield  {author} {\bibinfo {author} {\bibfnamefont {G.}~\bibnamefont
  {Knizia}}\ and\ \bibinfo {author} {\bibfnamefont {G.~K.-L.}\ \bibnamefont
  {Chan}},\ }\bibfield  {title} {\bibinfo {title} {Density matrix embedding: A
  simple alternative to dynamical mean-field theory},\ }\href
  {https://doi.org/10.1103/PhysRevLett.109.186404} {\bibfield  {journal}
  {\bibinfo  {journal} {Phys. Rev. Lett.}\ }\textbf {\bibinfo {volume} {109}},\
  \bibinfo {pages} {186404} (\bibinfo {year} {2012})}\BibitemShut {NoStop}%
\bibitem [{\citenamefont {Knizia}\ and\ \citenamefont
  {Chan}(2013)}]{Knizia2013}%
  \BibitemOpen
  \bibfield  {author} {\bibinfo {author} {\bibfnamefont {G.}~\bibnamefont
  {Knizia}}\ and\ \bibinfo {author} {\bibfnamefont {G.~K.-L.}\ \bibnamefont
  {Chan}},\ }\bibfield  {title} {\bibinfo {title} {Density matrix embedding: A
  strong-coupling quantum embedding theory},\ }\href@noop {} {\bibfield
  {journal} {\bibinfo  {journal} {Journal of chemical theory and computation}\
  }\textbf {\bibinfo {volume} {9}},\ \bibinfo {pages} {1428} (\bibinfo {year}
  {2013})}\BibitemShut {NoStop}%
\bibitem [{\citenamefont {Amaricci}\ \emph {et~al.}(2022)\citenamefont
  {Amaricci}, \citenamefont {Crippa}, \citenamefont {Scazzola}, \citenamefont
  {Petocchi}, \citenamefont {Mazza}, \citenamefont {{de Medici}},\ and\
  \citenamefont {Capone}}]{Amaricci2022}%
  \BibitemOpen
  \bibfield  {author} {\bibinfo {author} {\bibfnamefont {A.}~\bibnamefont
  {Amaricci}}, \bibinfo {author} {\bibfnamefont {L.}~\bibnamefont {Crippa}},
  \bibinfo {author} {\bibfnamefont {A.}~\bibnamefont {Scazzola}}, \bibinfo
  {author} {\bibfnamefont {F.}~\bibnamefont {Petocchi}}, \bibinfo {author}
  {\bibfnamefont {G.}~\bibnamefont {Mazza}}, \bibinfo {author} {\bibfnamefont
  {L.}~\bibnamefont {{de Medici}}},\ and\ \bibinfo {author} {\bibfnamefont
  {M.}~\bibnamefont {Capone}},\ }\bibfield  {title} {\bibinfo {title} {Edipack:
  A parallel exact diagonalization package for quantum impurity problems},\
  }\href {https://doi.org/https://doi.org/10.1016/j.cpc.2021.108261} {\bibfield
   {journal} {\bibinfo  {journal} {Computer Physics Communications}\ }\textbf
  {\bibinfo {volume} {273}},\ \bibinfo {pages} {108261} (\bibinfo {year}
  {2022})}\BibitemShut {NoStop}%
\bibitem [{\citenamefont {Zgid}\ and\ \citenamefont {Chan}(2011)}]{Zgid2011}%
  \BibitemOpen
  \bibfield  {author} {\bibinfo {author} {\bibfnamefont {D.}~\bibnamefont
  {Zgid}}\ and\ \bibinfo {author} {\bibfnamefont {G.~K.-L.}\ \bibnamefont
  {Chan}},\ }\bibfield  {title} {\bibinfo {title} {Dynamical mean-field theory
  from a quantum chemical perspective},\ }\href@noop {} {\bibfield  {journal}
  {\bibinfo  {journal} {J. Chem. Phys.}\ }\textbf {\bibinfo {volume} {134}},\
  \bibinfo {pages} {094115} (\bibinfo {year} {2011})}\BibitemShut {NoStop}%
\bibitem [{\citenamefont {Zgid}\ \emph {et~al.}(2012)\citenamefont {Zgid},
  \citenamefont {Gull},\ and\ \citenamefont {Chan}}]{Zgid2012}%
  \BibitemOpen
  \bibfield  {author} {\bibinfo {author} {\bibfnamefont {D.}~\bibnamefont
  {Zgid}}, \bibinfo {author} {\bibfnamefont {E.}~\bibnamefont {Gull}},\ and\
  \bibinfo {author} {\bibfnamefont {G.~K.-L.}\ \bibnamefont {Chan}},\
  }\bibfield  {title} {\bibinfo {title} {Truncated configuration interaction
  expansions as solvers for correlated quantum impurity models and dynamical
  mean-field theory},\ }\href@noop {} {\bibfield  {journal} {\bibinfo
  {journal} {Phys. Rev. B}\ }\textbf {\bibinfo {volume} {86}},\ \bibinfo
  {pages} {165128} (\bibinfo {year} {2012})}\BibitemShut {NoStop}%
\bibitem [{\citenamefont {Lu}\ \emph {et~al.}(2014)\citenamefont {Lu},
  \citenamefont {H\"oppner}, \citenamefont {Gunnarsson},\ and\ \citenamefont
  {Haverkort}}]{Lu2014}%
  \BibitemOpen
  \bibfield  {author} {\bibinfo {author} {\bibfnamefont {Y.}~\bibnamefont
  {Lu}}, \bibinfo {author} {\bibfnamefont {M.}~\bibnamefont {H\"oppner}},
  \bibinfo {author} {\bibfnamefont {O.}~\bibnamefont {Gunnarsson}},\ and\
  \bibinfo {author} {\bibfnamefont {M.~W.}\ \bibnamefont {Haverkort}},\
  }\bibfield  {title} {\bibinfo {title} {Efficient real-frequency solver for
  dynamical mean-field theory},\ }\href@noop {} {\bibfield  {journal} {\bibinfo
   {journal} {Phys. Rev. B}\ }\textbf {\bibinfo {volume} {90}},\ \bibinfo
  {pages} {085102} (\bibinfo {year} {2014})}\BibitemShut {NoStop}%
\bibitem [{\citenamefont {Go}\ and\ \citenamefont {Millis}(2015)}]{Go2015}%
  \BibitemOpen
  \bibfield  {author} {\bibinfo {author} {\bibfnamefont {A.}~\bibnamefont
  {Go}}\ and\ \bibinfo {author} {\bibfnamefont {A.~J.}\ \bibnamefont
  {Millis}},\ }\bibfield  {title} {\bibinfo {title} {Spatial correlations and
  the insulating phase of the high-t c cuprates: Insights from a
  configuration-interaction-based solver for dynamical mean field theory},\
  }\href@noop {} {\bibfield  {journal} {\bibinfo  {journal} {Phys. Rev. Lett.}\
  }\textbf {\bibinfo {volume} {114}},\ \bibinfo {pages} {016402} (\bibinfo
  {year} {2015})}\BibitemShut {NoStop}%
\bibitem [{\citenamefont {Go}\ and\ \citenamefont {Millis}(2017)}]{Go2017}%
  \BibitemOpen
  \bibfield  {author} {\bibinfo {author} {\bibfnamefont {A.}~\bibnamefont
  {Go}}\ and\ \bibinfo {author} {\bibfnamefont {A.~J.}\ \bibnamefont
  {Millis}},\ }\bibfield  {title} {\bibinfo {title} {Adaptively truncated
  hilbert space based impurity solver for dynamical mean-field theory},\
  }\href@noop {} {\bibfield  {journal} {\bibinfo  {journal} {Phys. Rev. B}\
  }\textbf {\bibinfo {volume} {96}},\ \bibinfo {pages} {085139} (\bibinfo
  {year} {2017})}\BibitemShut {NoStop}%
\bibitem [{\citenamefont {Mejuto-Zaera}\ \emph {et~al.}(2019)\citenamefont
  {Mejuto-Zaera}, \citenamefont {Tubman},\ and\ \citenamefont
  {Whaley}}]{Mejuto2019}%
  \BibitemOpen
  \bibfield  {author} {\bibinfo {author} {\bibfnamefont {C.}~\bibnamefont
  {Mejuto-Zaera}}, \bibinfo {author} {\bibfnamefont {N.~M.}\ \bibnamefont
  {Tubman}},\ and\ \bibinfo {author} {\bibfnamefont {K.~B.}\ \bibnamefont
  {Whaley}},\ }\bibfield  {title} {\bibinfo {title} {Dynamical mean field
  theory simulations with the adaptive sampling configuration interaction
  method},\ }\href@noop {} {\bibfield  {journal} {\bibinfo  {journal} {Physical
  Review B}\ }\textbf {\bibinfo {volume} {100}},\ \bibinfo {pages} {125165}
  (\bibinfo {year} {2019})}\BibitemShut {NoStop}%
\bibitem [{\citenamefont {Williams-Young}\ \emph {et~al.}(2023)\citenamefont
  {Williams-Young}, \citenamefont {Tubman}, \citenamefont {Mejuto-Zaera},\ and\
  \citenamefont {de~Jong}}]{WilliamsYoung2023}%
  \BibitemOpen
  \bibfield  {author} {\bibinfo {author} {\bibfnamefont {D.~B.}\ \bibnamefont
  {Williams-Young}}, \bibinfo {author} {\bibfnamefont {N.~M.}\ \bibnamefont
  {Tubman}}, \bibinfo {author} {\bibfnamefont {C.}~\bibnamefont
  {Mejuto-Zaera}},\ and\ \bibinfo {author} {\bibfnamefont {W.~A.}\ \bibnamefont
  {de~Jong}},\ }\href@noop {} {\bibinfo {title} {A parallel, distributed memory
  implementation of the adaptive sampling configuration interaction method}}
  (\bibinfo {year} {2023}),\ \Eprint {https://arxiv.org/abs/2303.05688}
  {arXiv:2303.05688 [physics.chem-ph]} \BibitemShut {NoStop}%
\bibitem [{\citenamefont {Werner}\ \emph {et~al.}(2023)\citenamefont {Werner},
  \citenamefont {Lotze},\ and\ \citenamefont {Arrigoni}}]{Werner2023}%
  \BibitemOpen
  \bibfield  {author} {\bibinfo {author} {\bibfnamefont {D.}~\bibnamefont
  {Werner}}, \bibinfo {author} {\bibfnamefont {J.}~\bibnamefont {Lotze}},\ and\
  \bibinfo {author} {\bibfnamefont {E.}~\bibnamefont {Arrigoni}},\ }\bibfield
  {title} {\bibinfo {title} {Configuration interaction based nonequilibrium
  steady state impurity solver},\ }\href
  {https://doi.org/10.1103/PhysRevB.107.075119} {\bibfield  {journal} {\bibinfo
   {journal} {Phys. Rev. B}\ }\textbf {\bibinfo {volume} {107}},\ \bibinfo
  {pages} {075119} (\bibinfo {year} {2023})}\BibitemShut {NoStop}%
\bibitem [{\citenamefont {Garcia}\ \emph {et~al.}(2004)\citenamefont {Garcia},
  \citenamefont {Hallberg},\ and\ \citenamefont {Rozenberg}}]{Garcia2004}%
  \BibitemOpen
  \bibfield  {author} {\bibinfo {author} {\bibfnamefont {D.~J.}\ \bibnamefont
  {Garcia}}, \bibinfo {author} {\bibfnamefont {K.}~\bibnamefont {Hallberg}},\
  and\ \bibinfo {author} {\bibfnamefont {M.~J.}\ \bibnamefont {Rozenberg}},\
  }\bibfield  {title} {\bibinfo {title} {Dynamical mean field theory with the
  density matrix renormalization group},\ }\href@noop {} {\bibfield  {journal}
  {\bibinfo  {journal} {Phys. Rev. Lett.}\ }\textbf {\bibinfo {volume} {93}},\
  \bibinfo {pages} {246403} (\bibinfo {year} {2004})}\BibitemShut {NoStop}%
\bibitem [{\citenamefont {Nishimoto}\ \emph {et~al.}(2004)\citenamefont
  {Nishimoto}, \citenamefont {Gebhard},\ and\ \citenamefont
  {Jeckelmann}}]{Nishimoto2004}%
  \BibitemOpen
  \bibfield  {author} {\bibinfo {author} {\bibfnamefont {S.}~\bibnamefont
  {Nishimoto}}, \bibinfo {author} {\bibfnamefont {F.}~\bibnamefont {Gebhard}},\
  and\ \bibinfo {author} {\bibfnamefont {E.}~\bibnamefont {Jeckelmann}},\
  }\bibfield  {title} {\bibinfo {title} {Dynamical density-matrix
  renormalization group for the mott–hubbard insulator in high dimensions},\
  }\href@noop {} {\bibfield  {journal} {\bibinfo  {journal} {J. Phys-Condens.
  Mat.}\ }\textbf {\bibinfo {volume} {16}},\ \bibinfo {pages} {7063} (\bibinfo
  {year} {2004})}\BibitemShut {NoStop}%
\bibitem [{\citenamefont {Peters}(2011)}]{Peters2011}%
  \BibitemOpen
  \bibfield  {author} {\bibinfo {author} {\bibfnamefont {R.}~\bibnamefont
  {Peters}},\ }\bibfield  {title} {\bibinfo {title} {Spectral functions for
  single- and multi-impurity models using density matrix renormalization
  group},\ }\href {https://doi.org/10.1103/PhysRevB.84.075139} {\bibfield
  {journal} {\bibinfo  {journal} {Phys. Rev. B}\ }\textbf {\bibinfo {volume}
  {84}},\ \bibinfo {pages} {075139} (\bibinfo {year} {2011})}\BibitemShut
  {NoStop}%
\bibitem [{\citenamefont {Wolf}\ \emph
  {et~al.}(2014{\natexlab{a}})\citenamefont {Wolf}, \citenamefont {McCulloch},\
  and\ \citenamefont {Schollw\"ock}}]{Wolf2014a}%
  \BibitemOpen
  \bibfield  {author} {\bibinfo {author} {\bibfnamefont {F.~A.}\ \bibnamefont
  {Wolf}}, \bibinfo {author} {\bibfnamefont {I.~P.}\ \bibnamefont
  {McCulloch}},\ and\ \bibinfo {author} {\bibfnamefont {U.}~\bibnamefont
  {Schollw\"ock}},\ }\bibfield  {title} {\bibinfo {title} {Solving
  nonequilibrium dynamical mean-field theory using matrix product states},\
  }\href@noop {} {\bibfield  {journal} {\bibinfo  {journal} {Phys. Rev. B}\
  }\textbf {\bibinfo {volume} {90}},\ \bibinfo {pages} {235131} (\bibinfo
  {year} {2014}{\natexlab{a}})}\BibitemShut {NoStop}%
\bibitem [{\citenamefont {Wolf}\ \emph
  {et~al.}(2014{\natexlab{b}})\citenamefont {Wolf}, \citenamefont {McCulloch},
  \citenamefont {Parcollet},\ and\ \citenamefont {Schollw{\"o}ck}}]{Wolf2014b}%
  \BibitemOpen
  \bibfield  {author} {\bibinfo {author} {\bibfnamefont {F.~A.}\ \bibnamefont
  {Wolf}}, \bibinfo {author} {\bibfnamefont {I.~P.}\ \bibnamefont {McCulloch}},
  \bibinfo {author} {\bibfnamefont {O.}~\bibnamefont {Parcollet}},\ and\
  \bibinfo {author} {\bibfnamefont {U.}~\bibnamefont {Schollw{\"o}ck}},\
  }\bibfield  {title} {\bibinfo {title} {Chebyshev matrix product state
  impurity solver for dynamical mean-field theory},\ }\href@noop {} {\bibfield
  {journal} {\bibinfo  {journal} {Physical Review B}\ }\textbf {\bibinfo
  {volume} {90}},\ \bibinfo {pages} {115124} (\bibinfo {year}
  {2014}{\natexlab{b}})}\BibitemShut {NoStop}%
\bibitem [{\citenamefont {Wolf}\ \emph
  {et~al.}(2015{\natexlab{a}})\citenamefont {Wolf}, \citenamefont {Go},
  \citenamefont {McCulloch}, \citenamefont {Millis},\ and\ \citenamefont
  {Schollw\"ock}}]{Wolf2015a}%
  \BibitemOpen
  \bibfield  {author} {\bibinfo {author} {\bibfnamefont {F.~A.}\ \bibnamefont
  {Wolf}}, \bibinfo {author} {\bibfnamefont {A.}~\bibnamefont {Go}}, \bibinfo
  {author} {\bibfnamefont {I.~P.}\ \bibnamefont {McCulloch}}, \bibinfo {author}
  {\bibfnamefont {A.~J.}\ \bibnamefont {Millis}},\ and\ \bibinfo {author}
  {\bibfnamefont {U.}~\bibnamefont {Schollw\"ock}},\ }\bibfield  {title}
  {\bibinfo {title} {Imaginary-time matrix product state impurity solver for
  dynamical mean-field theory},\ }\href@noop {} {\bibfield  {journal} {\bibinfo
   {journal} {Phys. Rev. X}\ }\textbf {\bibinfo {volume} {5}},\ \bibinfo
  {pages} {041032} (\bibinfo {year} {2015}{\natexlab{a}})}\BibitemShut
  {NoStop}%
\bibitem [{\citenamefont {Wolf}\ \emph
  {et~al.}(2015{\natexlab{b}})\citenamefont {Wolf}, \citenamefont {Justiniano},
  \citenamefont {McCulloch},\ and\ \citenamefont {Schollw{\"o}ck}}]{Wolf2015b}%
  \BibitemOpen
  \bibfield  {author} {\bibinfo {author} {\bibfnamefont {F.~A.}\ \bibnamefont
  {Wolf}}, \bibinfo {author} {\bibfnamefont {J.~A.}\ \bibnamefont
  {Justiniano}}, \bibinfo {author} {\bibfnamefont {I.~P.}\ \bibnamefont
  {McCulloch}},\ and\ \bibinfo {author} {\bibfnamefont {U.}~\bibnamefont
  {Schollw{\"o}ck}},\ }\bibfield  {title} {\bibinfo {title} {Spectral functions
  and time evolution from the chebyshev recursion},\ }\href@noop {} {\bibfield
  {journal} {\bibinfo  {journal} {Physical Review B}\ }\textbf {\bibinfo
  {volume} {91}},\ \bibinfo {pages} {115144} (\bibinfo {year}
  {2015}{\natexlab{b}})}\BibitemShut {NoStop}%
\bibitem [{\citenamefont {Bauernfeind}\ \emph {et~al.}(2017)\citenamefont
  {Bauernfeind}, \citenamefont {Zingl}, \citenamefont {Triebl}, \citenamefont
  {Aichhorn},\ and\ \citenamefont {Evertz}}]{Bauernfeind2017}%
  \BibitemOpen
  \bibfield  {author} {\bibinfo {author} {\bibfnamefont {D.}~\bibnamefont
  {Bauernfeind}}, \bibinfo {author} {\bibfnamefont {M.}~\bibnamefont {Zingl}},
  \bibinfo {author} {\bibfnamefont {R.}~\bibnamefont {Triebl}}, \bibinfo
  {author} {\bibfnamefont {M.}~\bibnamefont {Aichhorn}},\ and\ \bibinfo
  {author} {\bibfnamefont {H.~G.}\ \bibnamefont {Evertz}},\ }\bibfield  {title}
  {\bibinfo {title} {Fork tensor-product states: Efficient multiorbital
  real-time dmft solver},\ }\href {https://doi.org/10.1103/PhysRevX.7.031013}
  {\bibfield  {journal} {\bibinfo  {journal} {Phys. Rev. X}\ }\textbf {\bibinfo
  {volume} {7}},\ \bibinfo {pages} {031013} (\bibinfo {year}
  {2017})}\BibitemShut {NoStop}%
\bibitem [{\citenamefont {Paeckel}\ \emph {et~al.}(2019)\citenamefont
  {Paeckel}, \citenamefont {Köhler}, \citenamefont {Swoboda}, \citenamefont
  {Manmana}, \citenamefont {Schollwöck},\ and\ \citenamefont
  {Hubig}}]{Paeckel2019}%
  \BibitemOpen
  \bibfield  {author} {\bibinfo {author} {\bibfnamefont {S.}~\bibnamefont
  {Paeckel}}, \bibinfo {author} {\bibfnamefont {T.}~\bibnamefont {Köhler}},
  \bibinfo {author} {\bibfnamefont {A.}~\bibnamefont {Swoboda}}, \bibinfo
  {author} {\bibfnamefont {S.~R.}\ \bibnamefont {Manmana}}, \bibinfo {author}
  {\bibfnamefont {U.}~\bibnamefont {Schollwöck}},\ and\ \bibinfo {author}
  {\bibfnamefont {C.}~\bibnamefont {Hubig}},\ }\bibfield  {title} {\bibinfo
  {title} {Time-evolution methods for matrix-product states},\ }\href
  {https://doi.org/https://doi.org/10.1016/j.aop.2019.167998} {\bibfield
  {journal} {\bibinfo  {journal} {Annals of Physics}\ }\textbf {\bibinfo
  {volume} {411}},\ \bibinfo {pages} {167998} (\bibinfo {year}
  {2019})}\BibitemShut {NoStop}%
\bibitem [{\citenamefont {Mejuto-Zaera}\ \emph {et~al.}(2020)\citenamefont
  {Mejuto-Zaera}, \citenamefont {Zepeda-N{\'u}{\~n}ez}, \citenamefont
  {Lindsey}, \citenamefont {Tubman}, \citenamefont {Whaley},\ and\
  \citenamefont {Lin}}]{Mejuto2020}%
  \BibitemOpen
  \bibfield  {author} {\bibinfo {author} {\bibfnamefont {C.}~\bibnamefont
  {Mejuto-Zaera}}, \bibinfo {author} {\bibfnamefont {L.}~\bibnamefont
  {Zepeda-N{\'u}{\~n}ez}}, \bibinfo {author} {\bibfnamefont {M.}~\bibnamefont
  {Lindsey}}, \bibinfo {author} {\bibfnamefont {N.}~\bibnamefont {Tubman}},
  \bibinfo {author} {\bibfnamefont {B.}~\bibnamefont {Whaley}},\ and\ \bibinfo
  {author} {\bibfnamefont {L.}~\bibnamefont {Lin}},\ }\bibfield  {title}
  {\bibinfo {title} {Efficient hybridization fitting for dynamical mean-field
  theory via semi-definite relaxation},\ }\href@noop {} {\bibfield  {journal}
  {\bibinfo  {journal} {Physical Review B}\ }\textbf {\bibinfo {volume}
  {101}},\ \bibinfo {pages} {035143} (\bibinfo {year} {2020})}\BibitemShut
  {NoStop}%
\bibitem [{\citenamefont {de'Medici}\ \emph {et~al.}(2005)\citenamefont
  {de'Medici}, \citenamefont {Georges},\ and\ \citenamefont
  {Biermann}}]{deMedici2005}%
  \BibitemOpen
  \bibfield  {author} {\bibinfo {author} {\bibfnamefont {L.}~\bibnamefont
  {de'Medici}}, \bibinfo {author} {\bibfnamefont {A.}~\bibnamefont {Georges}},\
  and\ \bibinfo {author} {\bibfnamefont {S.}~\bibnamefont {Biermann}},\
  }\bibfield  {title} {\bibinfo {title} {Orbital-selective mott transition in
  multiband systems: Slave-spin representation and dynamical mean-field
  theory},\ }\href {https://doi.org/10.1103/PhysRevB.72.205124} {\bibfield
  {journal} {\bibinfo  {journal} {Phys. Rev. B}\ }\textbf {\bibinfo {volume}
  {72}},\ \bibinfo {pages} {205124} (\bibinfo {year} {2005})}\BibitemShut
  {NoStop}%
\bibitem [{\citenamefont {Ferrero}\ \emph {et~al.}(2005)\citenamefont
  {Ferrero}, \citenamefont {Becca}, \citenamefont {Fabrizio},\ and\
  \citenamefont {Capone}}]{Ferrero2005}%
  \BibitemOpen
  \bibfield  {author} {\bibinfo {author} {\bibfnamefont {M.}~\bibnamefont
  {Ferrero}}, \bibinfo {author} {\bibfnamefont {F.}~\bibnamefont {Becca}},
  \bibinfo {author} {\bibfnamefont {M.}~\bibnamefont {Fabrizio}},\ and\
  \bibinfo {author} {\bibfnamefont {M.}~\bibnamefont {Capone}},\ }\bibfield
  {title} {\bibinfo {title} {Dynamical behavior across the mott transition of
  two bands with different bandwidths},\ }\href@noop {} {\bibfield  {journal}
  {\bibinfo  {journal} {Phys. Rev. B}\ }\textbf {\bibinfo {volume} {72}},\
  \bibinfo {pages} {205126} (\bibinfo {year} {2005})}\BibitemShut {NoStop}%
\bibitem [{\citenamefont {Werner}\ and\ \citenamefont
  {Millis}(2007)}]{Werner2007}%
  \BibitemOpen
  \bibfield  {author} {\bibinfo {author} {\bibfnamefont {P.}~\bibnamefont
  {Werner}}\ and\ \bibinfo {author} {\bibfnamefont {A.~J.}\ \bibnamefont
  {Millis}},\ }\bibfield  {title} {\bibinfo {title} {High-spin to low-spin and
  orbital polarization transitions in multiorbital mott systems},\ }\href
  {https://doi.org/10.1103/PhysRevLett.99.126405} {\bibfield  {journal}
  {\bibinfo  {journal} {Phys. Rev. Lett.}\ }\textbf {\bibinfo {volume} {99}},\
  \bibinfo {pages} {126405} (\bibinfo {year} {2007})}\BibitemShut {NoStop}%
\bibitem [{\citenamefont {Isidori}\ \emph {et~al.}(2019)\citenamefont
  {Isidori}, \citenamefont {Berovi\ifmmode~\acute{c}\else \'{c}\fi{}},
  \citenamefont {Fanfarillo}, \citenamefont {de' Medici}, \citenamefont
  {Fabrizio},\ and\ \citenamefont {Capone}}]{Isidori2019}%
  \BibitemOpen
  \bibfield  {author} {\bibinfo {author} {\bibfnamefont {A.}~\bibnamefont
  {Isidori}}, \bibinfo {author} {\bibfnamefont {M.}~\bibnamefont
  {Berovi\ifmmode~\acute{c}\else \'{c}\fi{}}}, \bibinfo {author} {\bibfnamefont
  {L.}~\bibnamefont {Fanfarillo}}, \bibinfo {author} {\bibfnamefont
  {L.}~\bibnamefont {de' Medici}}, \bibinfo {author} {\bibfnamefont
  {M.}~\bibnamefont {Fabrizio}},\ and\ \bibinfo {author} {\bibfnamefont
  {M.}~\bibnamefont {Capone}},\ }\bibfield  {title} {\bibinfo {title} {Charge
  disproportionation, mixed valence, and janus effect in multiorbital systems:
  A tale of two insulators},\ }\href
  {https://doi.org/10.1103/PhysRevLett.122.186401} {\bibfield  {journal}
  {\bibinfo  {journal} {Phys. Rev. Lett.}\ }\textbf {\bibinfo {volume} {122}},\
  \bibinfo {pages} {186401} (\bibinfo {year} {2019})}\BibitemShut {NoStop}%
\bibitem [{\citenamefont {Imada}\ \emph {et~al.}(1998)\citenamefont {Imada},
  \citenamefont {Fujimori},\ and\ \citenamefont {Tokura}}]{Imada1998}%
  \BibitemOpen
  \bibfield  {author} {\bibinfo {author} {\bibfnamefont {M.}~\bibnamefont
  {Imada}}, \bibinfo {author} {\bibfnamefont {A.}~\bibnamefont {Fujimori}},\
  and\ \bibinfo {author} {\bibfnamefont {Y.}~\bibnamefont {Tokura}},\
  }\bibfield  {title} {\bibinfo {title} {Metal-insulator transitions},\
  }\href@noop {} {\bibfield  {journal} {\bibinfo  {journal} {Reviews of modern
  physics}\ }\textbf {\bibinfo {volume} {70}},\ \bibinfo {pages} {1039}
  (\bibinfo {year} {1998})}\BibitemShut {NoStop}%
\bibitem [{\citenamefont {Werner}\ \emph {et~al.}(2008)\citenamefont {Werner},
  \citenamefont {Gull}, \citenamefont {Troyer},\ and\ \citenamefont
  {Millis}}]{Werner2008}%
  \BibitemOpen
  \bibfield  {author} {\bibinfo {author} {\bibfnamefont {P.}~\bibnamefont
  {Werner}}, \bibinfo {author} {\bibfnamefont {E.}~\bibnamefont {Gull}},
  \bibinfo {author} {\bibfnamefont {M.}~\bibnamefont {Troyer}},\ and\ \bibinfo
  {author} {\bibfnamefont {A.~J.}\ \bibnamefont {Millis}},\ }\bibfield  {title}
  {\bibinfo {title} {Spin freezing transition and non-fermi-liquid self-energy
  in a three-orbital model},\ }\href
  {https://doi.org/10.1103/PhysRevLett.101.166405} {\bibfield  {journal}
  {\bibinfo  {journal} {Phys. Rev. Lett.}\ }\textbf {\bibinfo {volume} {101}},\
  \bibinfo {pages} {166405} (\bibinfo {year} {2008})}\BibitemShut {NoStop}%
\bibitem [{\citenamefont {Haule}\ and\ \citenamefont
  {Kotliar}(2009)}]{Haule2009}%
  \BibitemOpen
  \bibfield  {author} {\bibinfo {author} {\bibfnamefont {K.}~\bibnamefont
  {Haule}}\ and\ \bibinfo {author} {\bibfnamefont {G.}~\bibnamefont
  {Kotliar}},\ }\bibfield  {title} {\bibinfo {title} {Coherence–incoherence
  crossover in the normal state of iron oxypnictides and importance of hund's
  rule coupling},\ }\href {https://doi.org/10.1088/1367-2630/11/2/025021}
  {\bibfield  {journal} {\bibinfo  {journal} {New Journal of Physics}\ }\textbf
  {\bibinfo {volume} {11}},\ \bibinfo {pages} {025021} (\bibinfo {year}
  {2009})}\BibitemShut {NoStop}%
\bibitem [{\citenamefont {Hansmann}\ \emph {et~al.}(2010)\citenamefont
  {Hansmann}, \citenamefont {Arita}, \citenamefont {Toschi}, \citenamefont
  {Sakai}, \citenamefont {Sangiovanni},\ and\ \citenamefont
  {Held}}]{Hansmann2010}%
  \BibitemOpen
  \bibfield  {author} {\bibinfo {author} {\bibfnamefont {P.}~\bibnamefont
  {Hansmann}}, \bibinfo {author} {\bibfnamefont {R.}~\bibnamefont {Arita}},
  \bibinfo {author} {\bibfnamefont {A.}~\bibnamefont {Toschi}}, \bibinfo
  {author} {\bibfnamefont {S.}~\bibnamefont {Sakai}}, \bibinfo {author}
  {\bibfnamefont {G.}~\bibnamefont {Sangiovanni}},\ and\ \bibinfo {author}
  {\bibfnamefont {K.}~\bibnamefont {Held}},\ }\bibfield  {title} {\bibinfo
  {title} {Dichotomy between large local and small ordered magnetic moments in
  iron-based superconductors},\ }\href
  {https://doi.org/10.1103/PhysRevLett.104.197002} {\bibfield  {journal}
  {\bibinfo  {journal} {Phys. Rev. Lett.}\ }\textbf {\bibinfo {volume} {104}},\
  \bibinfo {pages} {197002} (\bibinfo {year} {2010})}\BibitemShut {NoStop}%
\bibitem [{\citenamefont {de' Medici}\ \emph {et~al.}(2014)\citenamefont {de'
  Medici}, \citenamefont {Giovannetti},\ and\ \citenamefont
  {Capone}}]{deMedici2014}%
  \BibitemOpen
  \bibfield  {author} {\bibinfo {author} {\bibfnamefont {L.}~\bibnamefont {de'
  Medici}}, \bibinfo {author} {\bibfnamefont {G.}~\bibnamefont {Giovannetti}},\
  and\ \bibinfo {author} {\bibfnamefont {M.}~\bibnamefont {Capone}},\
  }\bibfield  {title} {\bibinfo {title} {Selective mott physics as a key to
  iron superconductors},\ }\href
  {https://doi.org/10.1103/PhysRevLett.112.177001} {\bibfield  {journal}
  {\bibinfo  {journal} {Phys. Rev. Lett.}\ }\textbf {\bibinfo {volume} {112}},\
  \bibinfo {pages} {177001} (\bibinfo {year} {2014})}\BibitemShut {NoStop}%
\bibitem [{\citenamefont {Fanfarillo}\ and\ \citenamefont
  {Bascones}(2015)}]{Fanfarillo2015}%
  \BibitemOpen
  \bibfield  {author} {\bibinfo {author} {\bibfnamefont {L.}~\bibnamefont
  {Fanfarillo}}\ and\ \bibinfo {author} {\bibfnamefont {E.}~\bibnamefont
  {Bascones}},\ }\bibfield  {title} {\bibinfo {title} {Electronic correlations
  in hund metals},\ }\href {https://doi.org/10.1103/PhysRevB.92.075136}
  {\bibfield  {journal} {\bibinfo  {journal} {Phys. Rev. B}\ }\textbf {\bibinfo
  {volume} {92}},\ \bibinfo {pages} {075136} (\bibinfo {year}
  {2015})}\BibitemShut {NoStop}%
\bibitem [{\citenamefont {Hoshino}\ and\ \citenamefont
  {Werner}(2016)}]{Hoshino2016}%
  \BibitemOpen
  \bibfield  {author} {\bibinfo {author} {\bibfnamefont {S.}~\bibnamefont
  {Hoshino}}\ and\ \bibinfo {author} {\bibfnamefont {P.}~\bibnamefont
  {Werner}},\ }\bibfield  {title} {\bibinfo {title} {Electronic orders in
  multiorbital hubbard models with lifted orbital degeneracy},\ }\href
  {https://doi.org/10.1103/PhysRevB.93.155161} {\bibfield  {journal} {\bibinfo
  {journal} {Phys. Rev. B}\ }\textbf {\bibinfo {volume} {93}},\ \bibinfo
  {pages} {155161} (\bibinfo {year} {2016})}\BibitemShut {NoStop}%
\bibitem [{\citenamefont {Fanfarillo}\ \emph {et~al.}(2017)\citenamefont
  {Fanfarillo}, \citenamefont {Giovannetti}, \citenamefont {Capone},\ and\
  \citenamefont {Bascones}}]{Fanfarillo2017}%
  \BibitemOpen
  \bibfield  {author} {\bibinfo {author} {\bibfnamefont {L.}~\bibnamefont
  {Fanfarillo}}, \bibinfo {author} {\bibfnamefont {G.}~\bibnamefont
  {Giovannetti}}, \bibinfo {author} {\bibfnamefont {M.}~\bibnamefont
  {Capone}},\ and\ \bibinfo {author} {\bibfnamefont {E.}~\bibnamefont
  {Bascones}},\ }\bibfield  {title} {\bibinfo {title} {Nematicity at the hund's
  metal crossover in iron superconductors},\ }\href
  {https://doi.org/10.1103/PhysRevB.95.144511} {\bibfield  {journal} {\bibinfo
  {journal} {Phys. Rev. B}\ }\textbf {\bibinfo {volume} {95}},\ \bibinfo
  {pages} {144511} (\bibinfo {year} {2017})}\BibitemShut {NoStop}%
\bibitem [{\citenamefont {Villar~Arribi}\ and\ \citenamefont {de'
  Medici}(2018)}]{VillarArribi2018}%
  \BibitemOpen
  \bibfield  {author} {\bibinfo {author} {\bibfnamefont {P.}~\bibnamefont
  {Villar~Arribi}}\ and\ \bibinfo {author} {\bibfnamefont {L.}~\bibnamefont
  {de' Medici}},\ }\bibfield  {title} {\bibinfo {title} {{Hund-Enhanced
  Electronic Compressibility in FeSe and its Correlation with $T_c$}},\
  }\href@noop {} {\bibfield  {journal} {\bibinfo  {journal} {Phys. Rev. Lett.}\
  }\textbf {\bibinfo {volume} {121}},\ \bibinfo {pages} {197001} (\bibinfo
  {year} {2018})}\BibitemShut {NoStop}%
\bibitem [{\citenamefont {Villar~Arribi}\ and\ \citenamefont {de'
  Medici}(2021)}]{VillarArribi2021}%
  \BibitemOpen
  \bibfield  {author} {\bibinfo {author} {\bibfnamefont {P.}~\bibnamefont
  {Villar~Arribi}}\ and\ \bibinfo {author} {\bibfnamefont {L.}~\bibnamefont
  {de' Medici}},\ }\bibfield  {title} {\bibinfo {title} {Hund's metal crossover
  and superconductivity in the 111 family of iron-based superconductors},\
  }\href@noop {} {\bibfield  {journal} {\bibinfo  {journal} {Phys. Rev. B}\
  }\textbf {\bibinfo {volume} {104}},\ \bibinfo {pages} {125130} (\bibinfo
  {year} {2021})}\BibitemShut {NoStop}%
\bibitem [{\citenamefont {Anisimov}\ \emph {et~al.}(2002)\citenamefont
  {Anisimov}, \citenamefont {Nekrasov}, \citenamefont {Kondanov}, \citenamefont
  {Rice},\ and\ \citenamefont {Sigrist}}]{Anisimov2002}%
  \BibitemOpen
  \bibfield  {author} {\bibinfo {author} {\bibfnamefont {V.}~\bibnamefont
  {Anisimov}}, \bibinfo {author} {\bibfnamefont {I.}~\bibnamefont {Nekrasov}},
  \bibinfo {author} {\bibfnamefont {D.}~\bibnamefont {Kondanov}}, \bibinfo
  {author} {\bibfnamefont {T.}~\bibnamefont {Rice}},\ and\ \bibinfo {author}
  {\bibfnamefont {M.}~\bibnamefont {Sigrist}},\ }\bibfield  {title} {\bibinfo
  {title} {{Orbital-selective Mott-insulator in Ca$_{2-x}$Sr$_x$RuO$_4$}},\
  }\href@noop {} {\bibfield  {journal} {\bibinfo  {journal} {Eur. Phys. J. B}\
  }\textbf {\bibinfo {volume} {25}},\ \bibinfo {pages} {191} (\bibinfo {year}
  {2002})}\BibitemShut {NoStop}%
\bibitem [{\citenamefont {Liebsch}(2003)}]{Liebsch2003}%
  \BibitemOpen
  \bibfield  {author} {\bibinfo {author} {\bibfnamefont {A.}~\bibnamefont
  {Liebsch}},\ }\bibfield  {title} {\bibinfo {title} {Mott transitions in
  multiorbital systems},\ }\href@noop {} {\bibfield  {journal} {\bibinfo
  {journal} {Phys. Rev. Lett.}\ }\textbf {\bibinfo {volume} {91}},\ \bibinfo
  {pages} {226401} (\bibinfo {year} {2003})}\BibitemShut {NoStop}%
\bibitem [{\citenamefont {Koga}\ \emph {et~al.}(2004)\citenamefont {Koga},
  \citenamefont {Kawakami}, \citenamefont {Rice},\ and\ \citenamefont
  {Sigrist}}]{Koga2004}%
  \BibitemOpen
  \bibfield  {author} {\bibinfo {author} {\bibfnamefont {A.}~\bibnamefont
  {Koga}}, \bibinfo {author} {\bibfnamefont {N.}~\bibnamefont {Kawakami}},
  \bibinfo {author} {\bibfnamefont {T.~M.}\ \bibnamefont {Rice}},\ and\
  \bibinfo {author} {\bibfnamefont {M.}~\bibnamefont {Sigrist}},\ }\bibfield
  {title} {\bibinfo {title} {Orbital-selective mott transitions in the
  degenerate hubbard model},\ }\href@noop {} {\bibfield  {journal} {\bibinfo
  {journal} {Phys. Rev. Lett.}\ }\textbf {\bibinfo {volume} {92}},\ \bibinfo
  {pages} {216402} (\bibinfo {year} {2004})}\BibitemShut {NoStop}%
\bibitem [{\citenamefont {Fabrizio}\ \emph {et~al.}(1999)\citenamefont
  {Fabrizio}, \citenamefont {Gogolin},\ and\ \citenamefont
  {Nersesyan}}]{Fabrizio1999}%
  \BibitemOpen
  \bibfield  {author} {\bibinfo {author} {\bibfnamefont {M.}~\bibnamefont
  {Fabrizio}}, \bibinfo {author} {\bibfnamefont {A.~O.}\ \bibnamefont
  {Gogolin}},\ and\ \bibinfo {author} {\bibfnamefont {A.~A.}\ \bibnamefont
  {Nersesyan}},\ }\bibfield  {title} {\bibinfo {title} {From band insulator to
  mott insulator in one dimension},\ }\href@noop {} {\bibfield  {journal}
  {\bibinfo  {journal} {Phys. Rev. Lett.}\ }\textbf {\bibinfo {volume} {83}},\
  \bibinfo {pages} {2014} (\bibinfo {year} {1999})}\BibitemShut {NoStop}%
\bibitem [{\citenamefont {Nagaosa}\ and\ \citenamefont
  {Takimoto}(1986)}]{Nagaosa1986}%
  \BibitemOpen
  \bibfield  {author} {\bibinfo {author} {\bibfnamefont {N.}~\bibnamefont
  {Nagaosa}}\ and\ \bibinfo {author} {\bibfnamefont {J.}~\bibnamefont
  {Takimoto}},\ }\bibfield  {title} {\bibinfo {title} {Theory of neutral-ionic
  transition in organic crystals. i. monte carlo simulation of modified hubbard
  model},\ }\href@noop {} {\bibfield  {journal} {\bibinfo  {journal} {J. Phys.
  Soc. Jap.}\ }\textbf {\bibinfo {volume} {55}},\ \bibinfo {pages} {2735}
  (\bibinfo {year} {1986})}\BibitemShut {NoStop}%
\bibitem [{\citenamefont {Tincani}\ \emph {et~al.}(2009)\citenamefont
  {Tincani}, \citenamefont {Noack},\ and\ \citenamefont
  {Baeriswyl}}]{Tinkani2009}%
  \BibitemOpen
  \bibfield  {author} {\bibinfo {author} {\bibfnamefont {L.}~\bibnamefont
  {Tincani}}, \bibinfo {author} {\bibfnamefont {R.~M.}\ \bibnamefont {Noack}},\
  and\ \bibinfo {author} {\bibfnamefont {D.}~\bibnamefont {Baeriswyl}},\
  }\bibfield  {title} {\bibinfo {title} {Critical properties of the
  band-insulator-to-mott-insulator transition in the strong-coupling limit of
  the ionic hubbard model},\ }\href@noop {} {\bibfield  {journal} {\bibinfo
  {journal} {Phys. Rev. B}\ }\textbf {\bibinfo {volume} {79}},\ \bibinfo
  {pages} {165109} (\bibinfo {year} {2009})}\BibitemShut {NoStop}%
\bibitem [{Note1()}]{Note1}%
  \BibitemOpen
  \bibinfo {note} {It is worth noting that in this band insulator regime we
  encounter the problem of singular $\protect \sqrt {\Delta (\protect \mathbb
  {I}-\Delta )\protect \tmspace +\thickmuskip {.2777em}}$ in Eq.~\protect
  \eqref {eq:V_eq} and~\protect \eqref {eq:scf}. The band insulator phase can
  be nonetheless stabilized within the gGut approximation by adding a
  nearest-neighbor hopping coupling the lower and upper bands in the lattice
  Hamiltonian. This prevents the bands from completely filling up (emptying)
  while still allowing the insulating gap to open.}\BibitemShut {Stop}%
\bibitem [{\citenamefont {de' Medici}\ \emph {et~al.}(2011)\citenamefont {de'
  Medici}, \citenamefont {Mravlje},\ and\ \citenamefont
  {Georges}}]{deMedici2011}%
  \BibitemOpen
  \bibfield  {author} {\bibinfo {author} {\bibfnamefont {L.}~\bibnamefont {de'
  Medici}}, \bibinfo {author} {\bibfnamefont {J.}~\bibnamefont {Mravlje}},\
  and\ \bibinfo {author} {\bibfnamefont {A.}~\bibnamefont {Georges}},\
  }\bibfield  {title} {\bibinfo {title} {Janus-faced influence of hund's rule
  coupling in strongly correlated materials},\ }\href@noop {} {\bibfield
  {journal} {\bibinfo  {journal} {Phys. Rev. Lett.}\ }\textbf {\bibinfo
  {volume} {107}},\ \bibinfo {pages} {256401} (\bibinfo {year}
  {2011})}\BibitemShut {NoStop}%
\bibitem [{\citenamefont {Schir{\'o}}\ and\ \citenamefont
  {Fabrizio}(2010)}]{Schiro2010}%
  \BibitemOpen
  \bibfield  {author} {\bibinfo {author} {\bibfnamefont {M.}~\bibnamefont
  {Schir{\'o}}}\ and\ \bibinfo {author} {\bibfnamefont {M.}~\bibnamefont
  {Fabrizio}},\ }\bibfield  {title} {\bibinfo {title} {Time-dependent mean
  field theory for quench dynamics in correlated electron systems},\
  }\href@noop {} {\bibfield  {journal} {\bibinfo  {journal} {Phys. Rev. Lett.}\
  }\textbf {\bibinfo {volume} {105}},\ \bibinfo {pages} {076401} (\bibinfo
  {year} {2010})}\BibitemShut {NoStop}%
\bibitem [{\citenamefont {Sandri}\ and\ \citenamefont
  {Fabrizio}(2013)}]{Sandri2013}%
  \BibitemOpen
  \bibfield  {author} {\bibinfo {author} {\bibfnamefont {M.}~\bibnamefont
  {Sandri}}\ and\ \bibinfo {author} {\bibfnamefont {M.}~\bibnamefont
  {Fabrizio}},\ }\bibfield  {title} {\bibinfo {title} {Nonequilibrium dynamics
  in the antiferromagnetic hubbard model},\ }\href@noop {} {\bibfield
  {journal} {\bibinfo  {journal} {Phys. Rev. B}\ }\textbf {\bibinfo {volume}
  {88}},\ \bibinfo {pages} {165113} (\bibinfo {year} {2013})}\BibitemShut
  {NoStop}%
\bibitem [{\citenamefont {Fabrizio}(2013)}]{Fabrizio2013}%
  \BibitemOpen
  \bibfield  {author} {\bibinfo {author} {\bibfnamefont {M.}~\bibnamefont
  {Fabrizio}},\ }\bibfield  {title} {\bibinfo {title} {The out-of-equilibrium
  time-dependent gutzwiller approximation},\ }in\ \href@noop {} {\emph
  {\bibinfo {booktitle} {New Materials for Thermoelectric Applications: Theory
  and Experiment}}},\ \bibinfo {editor} {edited by\ \bibinfo {editor}
  {\bibfnamefont {V.}~\bibnamefont {Zlatic}}\ and\ \bibinfo {editor}
  {\bibfnamefont {A.}~\bibnamefont {Hewson}}}\ (\bibinfo  {publisher} {Springer
  Netherlands},\ \bibinfo {address} {Dordrecht},\ \bibinfo {year} {2013})\ pp.\
  \bibinfo {pages} {247--273}\BibitemShut {NoStop}%
\bibitem [{\citenamefont {Aoki}\ \emph {et~al.}(2014)\citenamefont {Aoki},
  \citenamefont {Tsuji}, \citenamefont {Eckstein}, \citenamefont {Kollar},
  \citenamefont {Oka},\ and\ \citenamefont {Werner}}]{Aoki2014}%
  \BibitemOpen
  \bibfield  {author} {\bibinfo {author} {\bibfnamefont {H.}~\bibnamefont
  {Aoki}}, \bibinfo {author} {\bibfnamefont {N.}~\bibnamefont {Tsuji}},
  \bibinfo {author} {\bibfnamefont {M.}~\bibnamefont {Eckstein}}, \bibinfo
  {author} {\bibfnamefont {M.}~\bibnamefont {Kollar}}, \bibinfo {author}
  {\bibfnamefont {T.}~\bibnamefont {Oka}},\ and\ \bibinfo {author}
  {\bibfnamefont {P.}~\bibnamefont {Werner}},\ }\bibfield  {title} {\bibinfo
  {title} {Nonequilibrium dynamical mean-field theory and its applications},\
  }\href {https://doi.org/10.1103/RevModPhys.86.779} {\bibfield  {journal}
  {\bibinfo  {journal} {Rev. Mod. Phys.}\ }\textbf {\bibinfo {volume} {86}},\
  \bibinfo {pages} {779} (\bibinfo {year} {2014})}\BibitemShut {NoStop}%
\bibitem [{\citenamefont {Guerci}(2019)}]{GuerciThesis}%
  \BibitemOpen
  \bibfield  {author} {\bibinfo {author} {\bibfnamefont {D.}~\bibnamefont
  {Guerci}},\ }\href {https://hdl.handle.net/20.500.11767/103994} {\emph
  {\bibinfo {title} {Beyond simple variational approaches to strongly
  correlated electron systems}}}\ (\bibinfo  {publisher} {PhD. Thesis, SISSA,
  Trieste, Italy},\ \bibinfo {year} {2019})\BibitemShut {NoStop}%
\bibitem [{\citenamefont {Guerci}\ \emph {et~al.}(2023)\citenamefont {Guerci},
  \citenamefont {Capone},\ and\ \citenamefont {Lanat\`a}}]{Guerci2023}%
  \BibitemOpen
  \bibfield  {author} {\bibinfo {author} {\bibfnamefont {D.}~\bibnamefont
  {Guerci}}, \bibinfo {author} {\bibfnamefont {M.}~\bibnamefont {Capone}},\
  and\ \bibinfo {author} {\bibfnamefont {N.}~\bibnamefont {Lanat\`a}},\
  }\bibfield  {title} {\bibinfo {title} {The time dependent ghost-gutzwiller
  dynamics},\ }\href@noop {} {\bibfield  {journal} {\bibinfo  {journal}
  {arXiv-preprint 2303.09584}\ } (\bibinfo {year} {2023})}\BibitemShut
  {NoStop}%
\bibitem [{\citenamefont {Lee}\ \emph {et~al.}(2023)\citenamefont {Lee},
  \citenamefont {Melnick}, \citenamefont {Adler}, \citenamefont {Lanatà},\
  and\ \citenamefont {Kotliar}}]{lee2023}%
  \BibitemOpen
  \bibfield  {author} {\bibinfo {author} {\bibfnamefont {T.-H.}\ \bibnamefont
  {Lee}}, \bibinfo {author} {\bibfnamefont {C.}~\bibnamefont {Melnick}},
  \bibinfo {author} {\bibfnamefont {R.}~\bibnamefont {Adler}}, \bibinfo
  {author} {\bibfnamefont {N.}~\bibnamefont {Lanatà}},\ and\ \bibinfo {author}
  {\bibfnamefont {G.}~\bibnamefont {Kotliar}},\ }\href@noop {} {\bibinfo
  {title} {Accuracy of ghost-rotationally-invariant slave-boson theory for
  multiorbital hubbard models and realistic materials}} (\bibinfo {year}
  {2023}),\ \Eprint {https://arxiv.org/abs/2305.11128} {arXiv:2305.11128
  [cond-mat.str-el]} \BibitemShut {NoStop}%
\bibitem [{\citenamefont {Lanatà}(2023)}]{lanata2023}%
  \BibitemOpen
  \bibfield  {author} {\bibinfo {author} {\bibfnamefont {N.}~\bibnamefont
  {Lanatà}},\ }\href@noop {} {\bibinfo {title} {Derivation of the ghost
  gutzwiller approximation from quantum embedding principles: the ghost density
  matrix embedding theory}} (\bibinfo {year} {2023}),\ \Eprint
  {https://arxiv.org/abs/2305.11895} {arXiv:2305.11895 [physics.comp-ph]}
  \BibitemShut {NoStop}%
\bibitem [{\citenamefont {Galassi}\ \emph {et~al.}(2021)\citenamefont {Galassi}
  \emph {et~al.}}]{gsl}%
  \BibitemOpen
  \bibfield  {author} {\bibinfo {author} {\bibfnamefont {M.}~\bibnamefont
  {Galassi}} \emph {et~al.},\ }\href {http://www.gnu.org/software/gsl/} {\emph
  {\bibinfo {title} {GNU Scientific Library Reference Manual (3rd Ed.)}}}\
  (\bibinfo {year} {2021})\BibitemShut {NoStop}%
\bibitem [{\citenamefont {Bertlmann}\ and\ \citenamefont
  {Krammer}(2008)}]{Bertlmann2008}%
  \BibitemOpen
  \bibfield  {author} {\bibinfo {author} {\bibfnamefont {R.}~\bibnamefont
  {Bertlmann}}\ and\ \bibinfo {author} {\bibfnamefont {P.}~\bibnamefont
  {Krammer}},\ }\bibfield  {title} {\bibinfo {title} {Bloch vectors for
  qudits},\ }\href@noop {} {\bibfield  {journal} {\bibinfo  {journal} {J. Phys.
  A: Math. Theor.}\ }\textbf {\bibinfo {volume} {41}},\ \bibinfo {pages}
  {235303} (\bibinfo {year} {2008})}\BibitemShut {NoStop}%
\end{thebibliography}
\end{document}